\documentclass[12pt,preprint]{aastex62}
\usepackage{natbib}
\usepackage{float}
\usepackage{subfloat}
\usepackage[caption=false]{subfig}

\pdfoutput=1

\newcommand\msunyr{M_{\odot}\,\rm yr^{-1}}

\def\micron{$\mu$m}

\begin{document}

\shortauthors{Muzerolle et al.}
\shorttitle{The DQ Tau inner disk}

\title{The Inner Disk and Accretion Flow of the Close Binary DQ Tau}

\author{James Muzerolle}
\affiliation{Space Telescope Science Institute, 3700 San Martin Dr., Baltimore MD 21218, USA}
\affiliation{Visiting Astronomer at the Infrared Telescope Facility, which is operated by the University of Hawaii under Cooperative Agreement no. NNX-08AE38A with the National Aeronautics and Space Administration, Science Mission Directorate, Planetary Astronomy Program.}
\author{Kevin Flaherty}
\affiliation{Department of Astronomy and Department of Physics, Williams College, Williamstown, MA 01267, USA}
\author{Zoltan Balog}
\affiliation{Center for Astronomy, University of Heidelberg, Heidelberg, Germany}
\author{Tracy Beck}
\affiliation{Space Telescope Science Institute, 3700 San Martin Dr., Baltimore, MD 21218, USA}
\author{Robert Gutermuth}
\affiliation{Department of Astronomy, University of Massachusetts, Amherst, MA 01003, USA}

\begin{abstract}
We present multi-epoch optical and near-infrared (NIR) photometry and
spectroscopy of the spectroscopic binary T Tauri star DQ Tau.
The photometric monitoring, obtained using SMARTS ANDICAM, recovers
the previously-seen correlation between optical flux and the 15.8-day
binary orbital period, with blue flux peaks occurring close to most
observed periastron passages.  For the first time, we find an even more
consistent correlation between orbital period and NIR brightness and color.
The onset of pulse events in the NIR on average precedes
those in the optical by a few days, with the rise usually starting
near apastron orbital phase.
We further obtained five epochs of spectroscopy
using IRTF SpeX, with a wavelength range of 0.8 to 5 microns, and derived
spectra of the infrared excess emission.
The shape and strength of the excess varies with time, with cooler
and weaker characteristic dust emission (T $\sim 1100-1300$ K) over
most of the binary orbit, and stronger/warmer dust emission
(T $\sim 1600$ K, indicative of dust sublimation) just before periastron
passage.  We suggest our results are broadly consistent with predictions
of simulations of disk structure and accretion flows around close binaries,
with the varying dust emission possibly tracing the evolution of accretion
streams falling inwards through a circumbinary disk cavity and feeding
the accretion pulses traced by the optical photometry and NIR emission lines.
However, our results also show more complicated behavior that is not fully
explained by this simple picture, and will require
further observations and modeling to fully interpret.
\end{abstract}


\section{Introduction}

While the disk accretion paradigm in young stars has tended to focus on
the simpler case of single stars, understanding how the process works in
binary systems is essential since most stars are multiple
\citep[e.g.][and references therein]{2013ARA&A..51..269D}.
Studies of young accreting binaries also provide
upper bounds on the dynamical effects of companions on disk
structure and evolution, and elucidate the potential for planet formation
in binary systems. Close (spectroscopic) binary systems are
particularly interesting since they are likely surrounded by circumbinary
accretion disks during the class I and II phases \citep{2000MNRAS.314...33B},
disks which are largely analogous to those around single stars
except within the central few AU.
Moreover, the relatively short periods of spectroscopic binaries,
typically ranging from days to weeks, enable multi-epoch monitoring
campaigns capable of characterizing variability in the accretion and
inner disk emission that may trace dynamically-induced effects on
the material closest to the stars.

The spectroscopic binary T Tauri star DQ Tau has been a target of
particular interest.  \citet{1997AJ....113.1841M} first characterized its
orbital parameters ($P=15.8$ days, $e=0.56$, $a \sim$ 0.14 AU),
and also discovered
a correlation between orbital phase and optical photometric variability,
in which the light curves exhibited sharp increases in brightness
at or just before periastron passages.
A similar correlation of spectroscopic signatures
of accretion activity such as H$\alpha$ emission and continuum veiling
was also found by \citet{1997AJ....114..781B}.  These results suggested that
the accretion flow onto the stars was highly modulated by their
orbital motion, repeatedly peaking in intensity when they drew close
to each other.  Such behavior was predicted by hydrodynamical simulations
of circumbinary disk accretion, which showed that torques generated
by the binary motion create a low-density cavity in the disk out to a distance
of $\sim 2.5a$; accretion then precedes onto the stars via dynamical streams
of denser material that are
repeatedly torn off the inner edge of the disk near apastron orbital
phase and reach a maximum flow of gas near periastron phase, before
being disrupted as the stars move further apart \citep{1996ApJ...467L..77A}.
This process has come to be known as ``pulsed" accretion.
Subsequent simulations of circumbinary disks around both young stars and
compact objects, with increasingly sophisticated computational techniques
and exploration of parameter space, have shown the same general features,
and elucidated the effects of different binary orbital architectures
on the strength and periodicity of the accretion
\citep{2002A&A...387..550G, 2011MNRAS.413.2679D, 2012ApJ...749..118S,
2015MNRAS.448.3545D, 2016ApJ...827...43M}.

Subsequent observations of DQ Tau have further clarified its behavior
and revealed more complexity.  \citet{2001ApJ...551..454C} detected CO
fundamental emission, constraining the location of the emitting gas
to be within the putative circumbinary disk hole.  \citet{2009ApJ...696L.111B}
resolved K-band emission using interferometry, also locating
the emission region inside of the expected disk inner edge.
Observations of flares at X-ray and millimeter wavelengths
\citep{2008A&A...492L..21S, 2010A&A...521A..32S, 2011ApJ...730....6G}
found evidence of enhanced activity near periastron, suggestive of
magnetic reconnection events induced as the magnetospheres of
the two stars interact/collide. \citet[hereafter, Bary14]{2014ApJ...792...64B}
recovered the orbital phase dependence of accretion using Pa$\beta$
emission from multi-epoch near infrared spectroscopy; however, they also
discovered a surprising increase in accretion luminosity near {\it apastron}
phase at one epoch.  More recent optical photometric investigations
have also uncovered complex light curve behavior, with occasional
brightenings not associated with periastron
\citep{2017ApJ...835....8T, 2018ApJ...862...44K},
although a correlation with orbital phase
remains the dominant feature.

The signature of pulsed accretion remains extremely rare.
Only one other young close binary system has been found that exhibits
unambiguous brightenings associated with periastron passages
\citep[TWA3;][]{2017ApJ...842L..12T}.  \citet{2007AJ....134..241J} found
a periodicity in the optical light curve of UZ Tau E which matched
the orbital period, however the flux variations were much slower with
low amplitude, and direct accretion probes such as H$\alpha$ showed
inconclusive correlations.  Three protostellar objects have also been
found to exhibit periodic infrared brightenings that look very much
like pulsed accretion
\citep{2013Natur.493..378M, 2015ApJ...813..107H, 2016ApJ...833..104F},
although no evidence of binarity has yet been published.

Other aspects of the expected circumbinary disk structure have
been seen in wider binaries where features can be more readily
spatially resolved, such as circumbinary rings and streams
\citep[e.g., GG Tau;][]{2016A&ARv..24....5D}.
In close binaries such as DQ Tau, these regions
of the system can be probed with multi-epoch infrared observations,
particularly at wavelengths $\ge 2 \mu$m where warm dust emission
becomes significant.  Infrared photometry in particular has been
lacking, however; we thus initiated a multi-year campaign to obtain
near-infrared (NIR) photometric monitoring, with simultaneous optical
observations, spanning multiple orbital cycles.
In this paper, we present the results of the first two
seasons of photometry, along with contemporaneous
0.8-5 $\mu$m spectroscopy at a more limited set of epochs.
In section 2, we describe the observations and data reduction.
Section 3 details the photometric analysis of light curves,
color-color comparisons, and periodicities and timing, as well as
the spectroscopic analysis of spectral typing, veiling measurements,
and derivation of excess spectra.  Finally, we discuss our results
in the context of the pulsed accretion model in section 4.

For the purposes of our various analyses, we adopted the most recent
and robust determinations of the DQ Tau system parameters from
the literature, as listed in Table~\ref{params}.  This includes updating
the distance from the canonical Taurus region value of 140 pc
to the specific value of 196 pc recently measured by Gaia
\citep{2018AJ....156...58B}.  Despite the possibility of systematic
errors due to the binarity of the system, we believe this new larger value
is likely correct; DQ Tau is part of a small subgroup of five stars that lie
within the L1558 cloud southeast of the main concentration of stars
in the Taurus star forming region; their median Gaia distance
is 196 pc \citep{2018AJ....156..271L}.

\begin{deluxetable}{lcc}
\tablewidth{0pt}
\tablecaption{Literature properties of DQ Tau}
\tablehead{
\colhead{property} &
\colhead{adopted value} &
\colhead{reference}}
\startdata
$P$ (days) & 15.80158 & \citet{2016ApJ...818..156C}\\
$e$ & 0.568 & \citet{2016ApJ...818..156C}\\
$T_{peri}$ (HJD-2,400,000) & 47433.507 & \citet{2016ApJ...818..156C}\\
d (pc) & 196 & \citet{2018AJ....156...58B}\tablenotemark{a}\\
$T_{eff}$ (K) & 3700 & \citet{2016ApJ...818..156C}\\
$i$ (deg) & 158 & \citet{2016ApJ...818..156C}\\
$M_1 + M_2$ ($M_{\odot}$) & 1.21 & \citet{2016ApJ...818..156C}\tablenotemark{b}\\
$L_1 + L_2$ ($L_{\odot}$) & 0.64 & \citet{2016ApJ...818..156C}\tablenotemark{c}\\
$a$ (AU) & 0.13 & \citet{2016ApJ...818..156C}\\
$R_{co}$ (AU) & 0.034 & \citet{2018ApJ...862...44K}\tablenotemark{d}\\
\enddata
\tablenotetext{a}{Based on the Gaia DR2 catalog.}
\tablenotetext{b}{Estimated by Czekala et al. assuming a distance of 140 pc.}
\tablenotetext{c}{Scaled up by a factor of two to account for the Gaia distance of 196 pc.}
\tablenotetext{d}{Corotation radius calculated given the rotation period
measured by K\'osp\'al et al. (3.017 days), and the stellar mass (not adjusted for
the difference in distance, and assuming
both stars are identical).}
\label{params}
\end{deluxetable}

\section{Observations}

\subsection{SMARTS Photometry}
We observed DQ Tau with the ANDICAM instrument on the CTIO 1.3m telescope,
operated by the SMARTS Consortium.  ANDICAM is a dual-channel imager that
enables simultaneous observations at two band passes in the optical and
near-infrared.  The optical channel has a field of view is $\sim 6$' x 6' and
detector pixel scale of $\sim 0.3$ arcsec, while the NIR channel has a field
of view of $\sim 2.4$' x 2.4' and detector pixel scale of $\sim 0.2$ arcsec.
We used $BVI$ filters (standard KPNO
Johnson-Cousins) with the optical channel and $JHK$ filters (CIT/CTIO) with
 the infrared channel.
Table~\ref{obs} summarizes the exposure times and observation date ranges.
The NIR channel has an internal tip-tilt mirror to enable small-scale
dithering; six exposures were taken at each filter,
with each exposure separated by a small dither offset of $\sim 20$ arcseconds.
The observations were obtained over two seasons from fall 2012 to winter 2014;
each season had almost continuous nightly coverage (excluding bad
weather or scheduling pre-emptions) over two- to four-month periods,
during which time the star was observable below 2 airmasses.
The total sequence of exposures taken with 3 pairs of filters
each night took about 10 minutes to execute.

\begin{deluxetable}{lccccccc}
\tabletypesize{\small}
\tablewidth{0pt}
\tablecaption{SMARTS DQ Tau observation summary}
\tablehead{
\colhead{date range} &
\multicolumn{6}{c}{exposures\tablenotemark{a}} &
\colhead{N$_{obs}$}\\
\colhead{} &
\colhead{B} &
\colhead{V} &
\colhead{I} &
\colhead{J} &
\colhead{H} &
\colhead{K} &
\colhead{}}
\startdata
Nov. 15 2012 - Jan. 30 2013 & 3x30 & 3x15 & 3x10 & 6x12 & 6x6 & 6x7 & 62\\
Sep. 9 2013 - Feb. 3 2014 & 3x30 & 3x15 & 3x15 & 6x15 & 6x9 & 6x9 & 103
\enddata
\tablenotetext{a}{Number of exposures and exposure time in seconds,
per filter and observation.}
\label{obs}
\end{deluxetable}

The optical data are automatically processed by the SMARTS pipeline,
including bias and zero subtraction and flat field correction.  We measured
stellar photometry using the resulting archived products, as described below.
The NIR images are not automatically processed, other than having a 2x2
pixel binning applied.  For each set of dithered exposures at each filter,
we created a median sky image and subtracted from each exposure.  We then
divided the sky-subtracted images by flat fields constructed from dome flat
exposures at each filter.

We measured stellar photometry using the archived pipeline products in
the optical and the sky-subtracted, flat fielded images in the NIR.
This was done with aperture photometry, using an aperture radius of
20/15 pixels and sky annulus of 25-32/30-35 pixels in the optical/NIR.
Suitably bright comparison stars within the field provided relative photometry,
calibrating out variable nightly weather conditions; three were used in
the optical images and one in the NIR.  Most of these comparison stars
are fainter than DQ Tau at all bands, and thus are the limiting factor
in the final photometric uncertainties in most cases.  In poor weather
conditions, the NIR sky background was variable on timescales of
minutes or less, which sometimes contributed the largest source of
photometric error at the $JHK$ bands.  Relative magnitudes were computed for
each exposure in each band, and the final results averaged over all exposures
(and all comparison stars in the optical) for a given band on each night,
including one iteration of outlier rejection, with the standard deviation
of these values taken as the uncertainty.  By cross-checking the optical
comparison stars, we demonstrated excellent repeatability (and ruled out any
significant intrinsic variability) with an overall precision of $\sim 0.04$,
0.02, and 0.015 magnitudes at $B$, $V$, and $I$.  The relative precision
of the NIR photometry, as estimated by comparing the primary comparison star
with a fainter third star in the field, is $\sim 0.01$, 0.02, and 0.04
magnitudes at $J$, $H$, and $K$ (the fainter star is bluer, so these
relative measurements are more uncertain at longer wavelengths).

To convert the optical photometry to the Johnson-Cousins system,
we calibrated the three comparison stars using contemporaneous standard star
observations taken during the 2012 season on ostensibly photometric nights.
A total of 35 observations of the Landolt standard TPhe D were taken on
the same nights as our DQ Tau observations.  The zero points and
color terms for these are given on the SMARTS consortium website, and we
used them to convert the instrumental magnitudes to the standard system.
Table~\ref{compstars} gives the resultant magnitudes.
The extinction coefficients are the largest source of uncertainty in this
conversion because of the limited number of measurements as a function
of airmass; we used the ``default" values provided on the SMARTS website.
We conservatively estimate overall absolute uncertainties of about
0.2 magnitudes in each optical band.  $B$ and $V$ photometry
for all three stars are provided in the UCAC4 catalog \citep{2013AJ....145...44Z},
and are within 0.1 magnitudes of our values.  \citet{2010A&A...521A..32S} also
derived photometry for all three stars at $V$ and $I$ that agree to within
0.2 magnitudes.  In the NIR, we used photometry of the primary comparison
star from 2MASS to convert the relative magnitudes of DQ Tau directly
to the CIT system \citep{2001AJ....121.2851C}; the absolute accuracy is then
limited by the
uncertainties of the 2MASS measurements and the transformations
between the 2MASS and CIT systems (combined, about 0.05 magnitudes
in each band).  Table~\ref{photometry} shows a truncated set of the final
calibrated photometry (the full version will be made available online).

\begin{deluxetable}{lccrrrrrr}
\tabletypesize{\small}
\tablewidth{0pt}
\tablecaption{Comparison star photometry}
\tablehead{
\colhead{ID} &
\colhead{RA\tablenotemark{a}} &
\colhead{DEC\tablenotemark{a}} &
\colhead{B} &
\colhead{V} &
\colhead{I} &
\colhead{J\tablenotemark{a}} &
\colhead{H\tablenotemark{a}} &
\colhead{K\tablenotemark{a}}}
\startdata
1\tablenotemark{b}  &  04:46:40.79 & +16:57:50.4 & 15.19 & 13.56 & 11.66 & 10.27 & 9.55 & 9.32\\
2\tablenotemark{b}  &  04:46:39.64 &  +17:00:04.3 & 15.84 & 14.38 & 12.46 & 10.98 & 10.43 & 10.16\\
3  &  04:46:46.15 &  +17:00:29.1 & 16.02 & 14.79 & 13.22 & 12.02 & 11.51 & 11.33
\enddata
\tablenotetext{a}{Coordinates and magnitudes from 2MASS.}
\tablenotetext{b}{Not in ANDICAM NIR field of view.}
\label{compstars}
\end{deluxetable}

\begin{deluxetable}{lcccccccccccc}
\tabletypesize{\small}
\tablewidth{0pt}
\tablecaption{SMARTS/ANDICAM photometry for DQ Tau}
\tablehead{
\colhead{JD -2450000} &
\colhead{B} &
\colhead{$\sigma_B$} & 
\colhead{V} &
\colhead{$\sigma_V$} &
\colhead{I} &
\colhead{$\sigma_I$} &
\colhead{J} &
\colhead{$\sigma_J$} &
\colhead{H} &
\colhead{$\sigma_H$} &
\colhead{K} &
\colhead{$\sigma_K$}}
\startdata
6246.72 & 15.0582 &  0.0120 & 13.4431 &  0.0160 & 11.1853 &  0.0256 &  9.4064 &  0.0074 &  8.4451 &  0.0031 &  7.9183 &  0.0085 \\
6247.65 & 15.0151 &  0.0242 & 13.4501 &  0.0104 & 11.1837 &  0.0094 &  9.4013 &  0.0083 &  8.4310 &  0.0122 &  7.8902 &  0.0004 \\
6248.63 & 14.6287 &  0.0201 & 13.1966 &  0.0120 & 11.0268 &  0.0112 &  9.3191 &  0.0017 &  8.3424 &  0.0267 &  7.7732 &  0.0135 \\
6249.66 & 13.6157 &  0.0153 & 12.4896 &  0.0093 & 10.5984 &  0.0152 &  9.0337 &  0.0179 &  8.0772 &  0.0062 &  7.4715 &  0.0087 \\
6250.63 & 14.4273 &  0.0237 & 13.1242 &  0.0165 & 10.9752 &  0.0241 &  9.1916 &  0.0145 &  8.2149 &  0.0033 &  7.6565 &  0.0192 \\
6251.65 & 14.6657 &  0.0189 & 13.2381 &  0.0104 & 11.0688 &  0.0107 &  9.3150 &  0.0067 &  8.3561 &  0.0066 &  7.8098 &  0.0078 \\
6252.62 & 14.8033 &  0.0331 & 13.2841 &  0.0188 & 11.0839 &  0.0126 &  9.3405 &  0.0126 &  8.3890 &  0.0088 &  7.8629 &  0.0104 \\
6253.67 & 15.1059 &  0.0312 & 13.4964 &  0.0148 & 11.2090 &  0.0099 &  9.4460 &  0.0035 &  8.4848 &  0.0032 &  7.9601 &  0.0048 \\
6254.64 & 15.0925 &  0.0174 & 13.4748 &  0.0049 & 11.1824 &  0.0127 &  9.4246 &  0.0050 &  8.4928 &  0.0047 &  7.9534 &  0.0133 \\
6255.68 & 15.0436 &  0.0370 & 13.4693 &  0.0187 & 11.1732 &  0.0074 &  9.4282 &  0.0092 &  8.4614 &  0.0035 &  7.9331 &  0.0101
\enddata
\tablecomments{The quoted uncertainties include only random measurement errors.  Values of -999 indicate missing data.}
\label{photometry}
\end{deluxetable}

\subsection{SpeX Spectroscopy}

\begin{figure}[H]
\includegraphics[scale=0.8]{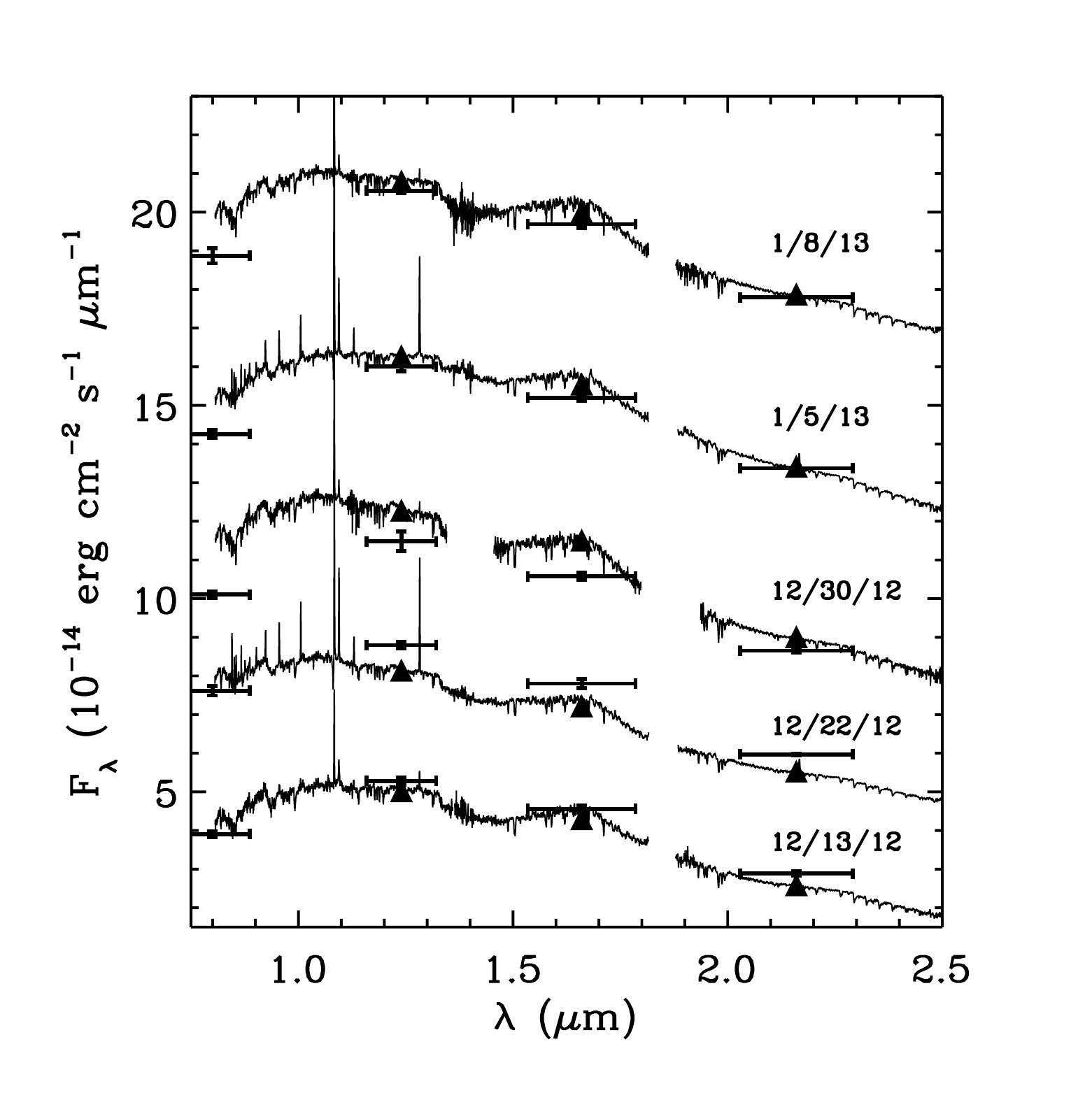}
\caption{Comparison of the NIR portion of the SpeX spectra (solid lines) for
all epochs with the ANDICAM photometric measurements nearest in time (error
bars; the horizontal portion represents the ANDICAM bandpasses).
For clarity, each set of data after the 12/13/2012 set have been shifted upward
by the following amounts in order of increasing time: 3, 7, 11, 16.
Solid triangles indicate the spectral flux convolved by the appropriate
ANDICAM bandpass at the J, H, and K bands.
The absolute flux levels of the spectra differ from the ANDICAM photometry
by $\sim15$\% or less, and the variations between bandpasses at each epoch
are within $\sim3$\%.
\label{compspec}}
\end{figure}

We observed DQ Tau with the SpeX spectrograph \citep{2003PASP..115..362R}
at IRTF on December 13, 22, and 30 2012 and January 5 and 8 2013.
All observations used both SXD and
LXD modes, for a combined wavelength coverage of 0.8 to 5 \micron,
and a slit width of 0.8", for a spectral resolution of $\sim 1000$.
The position angle was adjusted to keep the slit at the parallactic angle
during each observation in order to minimize slit losses from atmospheric
refraction.  Total exposure times were typically 6 to 8 minutes for
SXD mode and 30 to 40 minutes for LXD mode, all split into multiple nods
in the typical ABBA pattern for background subtraction. 
The data were reduced and spectra extracted using the Spextool package
\citep{2004PASP..116..362C}, which includes routines for sky subtraction,
flat fielding, tracing and extraction of each spectral order,
telluric correction, and order matching and combination.  The telluric
correction step was done using spectra of A0 stars observed near in time
and airmass to each DQ Tau observation, combining with a model spectrum
of Vega in order to correct for photospheric absorption lines
\citep{2003PASP..115..389V}.  The final spectra have a very accurate spectral shape,
although the absolute flux level (estimated by extrapolating from
the optical brightness of the telluric standard) can have larger errors
depending on the weather conditions (Fig.~\ref{compspec}).

\section{Results}

\subsection{Photometric Behavior}

\begin{figure}[H]
\plotone{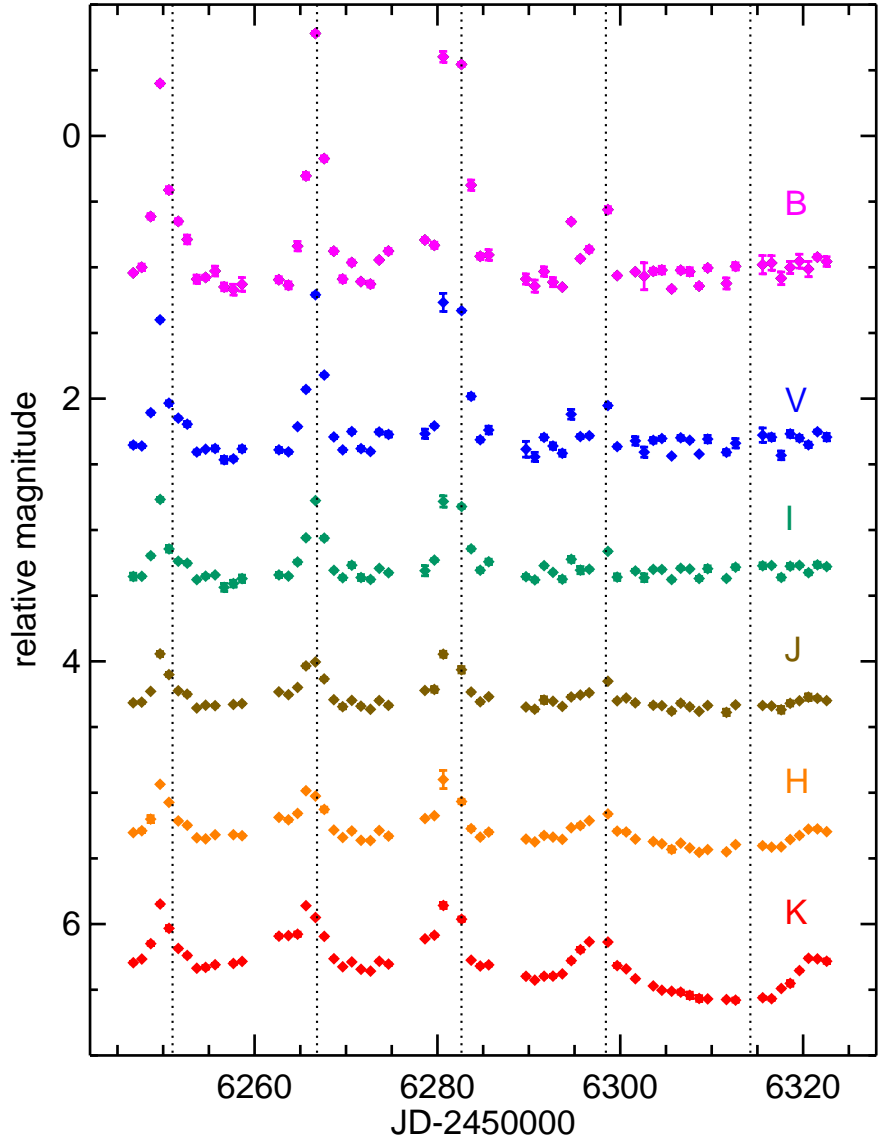}
\caption{SMARTS/ANDICAM BVIJHK light curves of DQ Tau, in relative magnitudes,
for the 2012-2013 season.
The epochs of periastron passage are marked with dotted black lines.
For clarity, each band has been median-subtracted and offset by an arbitrary amount.
\label{smphot12}}
\end{figure}

\begin{figure}[H]
\plotone{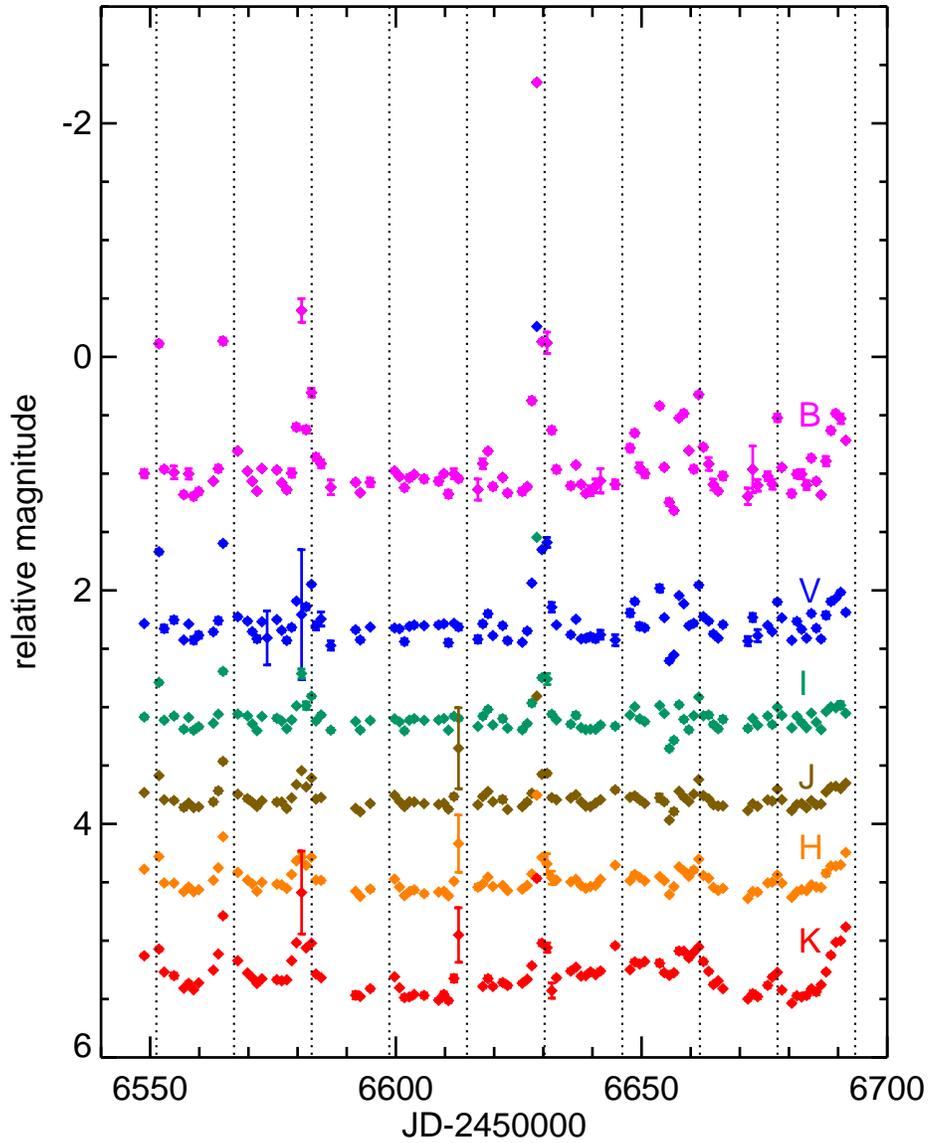}
\caption{Same as in Figure~\ref{smphot12}, for the 2013-2014 season.
\label{smphot13}}
\end{figure}

The time series photometry for the 2012-2013 and 2013-2014 seasons are
shown in Figures~\ref{smphot12} and~\ref{smphot13}, respectively.  Repeated flux
increases (hereafter referred to as ``pulses") above a flat or slowly varying
baseline level are readily apparent at all bands in both the optical and NIR.
Pulse events occur within most of the binary orbits covered by our data,
and usually (but not always) peak at or just before the time of periastron passage.
The pulses are typically sharply peaked and have durations ranging from
$\sim$2 to 6 days.  Of the 14 binary orbits fully covered over both seasons,
only two appear to lack obvious pulse events in the optical (the ones with
periastron passages near JD 2456315 and 2456600); however, because of gaps
in coverage due to bad weather, we cannot rule out the presence of
very short pulses with durations of 1-2 days in these cases.
In the NIR, only one orbit shows
no related pulse (JD 2456315), although the timing of the increase at the
very end of the 2012-2013 monitoring leads to ambiguity as to whether it was
associated with the preceding orbit or the next (mostly uncovered) one.
The timing of the pulse peaks in many cases appears to be wavelength-dependent,
with the peaks at $BVIJ$ typically occurring at the same time, while some peaks
at $H$ and $K$ occur up to 1 to 2 days beforehand.
The amplitude of the optical peaks are strongly variable, with increases above
the baseline at $B$ mostly ranging from $\sim 0.5$ to 2 magnitudes,
and up to 3.5 magnitudes in one extraordinary case (near JD 2456630).
In the NIR, the amplitudes are smaller and somewhat more regular,
with most pulses at about 0.5 magnitudes at $K$.

\begin{figure}[H]
\includegraphics[scale=0.9]{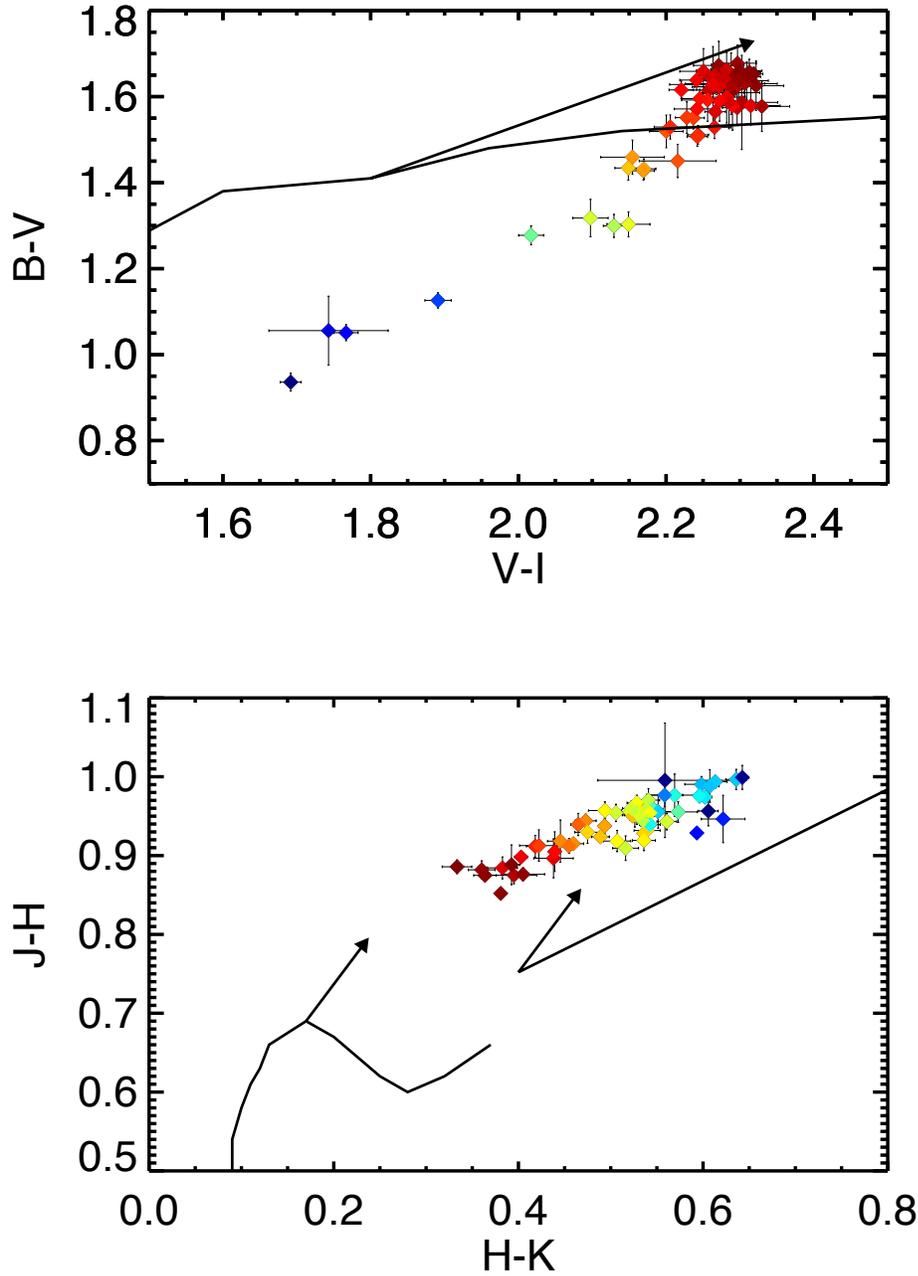}
\caption{SMARTS/ANDICAM optical and NIR color-color diagrams for
the 2012-2013 season.  The symbol color scheme scales with B (upper panel)
and K (lower panel) magnitudes,
where indigo is the brightest observation and dark red is the faintest.
The dwarf star color sequences \citep[using colors from][]{1995ApJS..101..117K}
are shown with the black solid lines; the arrows depict reddening vectors
for an M0 photosphere with $A_V=1$. The CTTS locus \citep{1997AJ....114..288M}
is also shown in the lower panel, with a reddening vector at its blue end.
\label{smcc12}}
\end{figure}

\begin{figure}[H]
\includegraphics[scale=0.9]{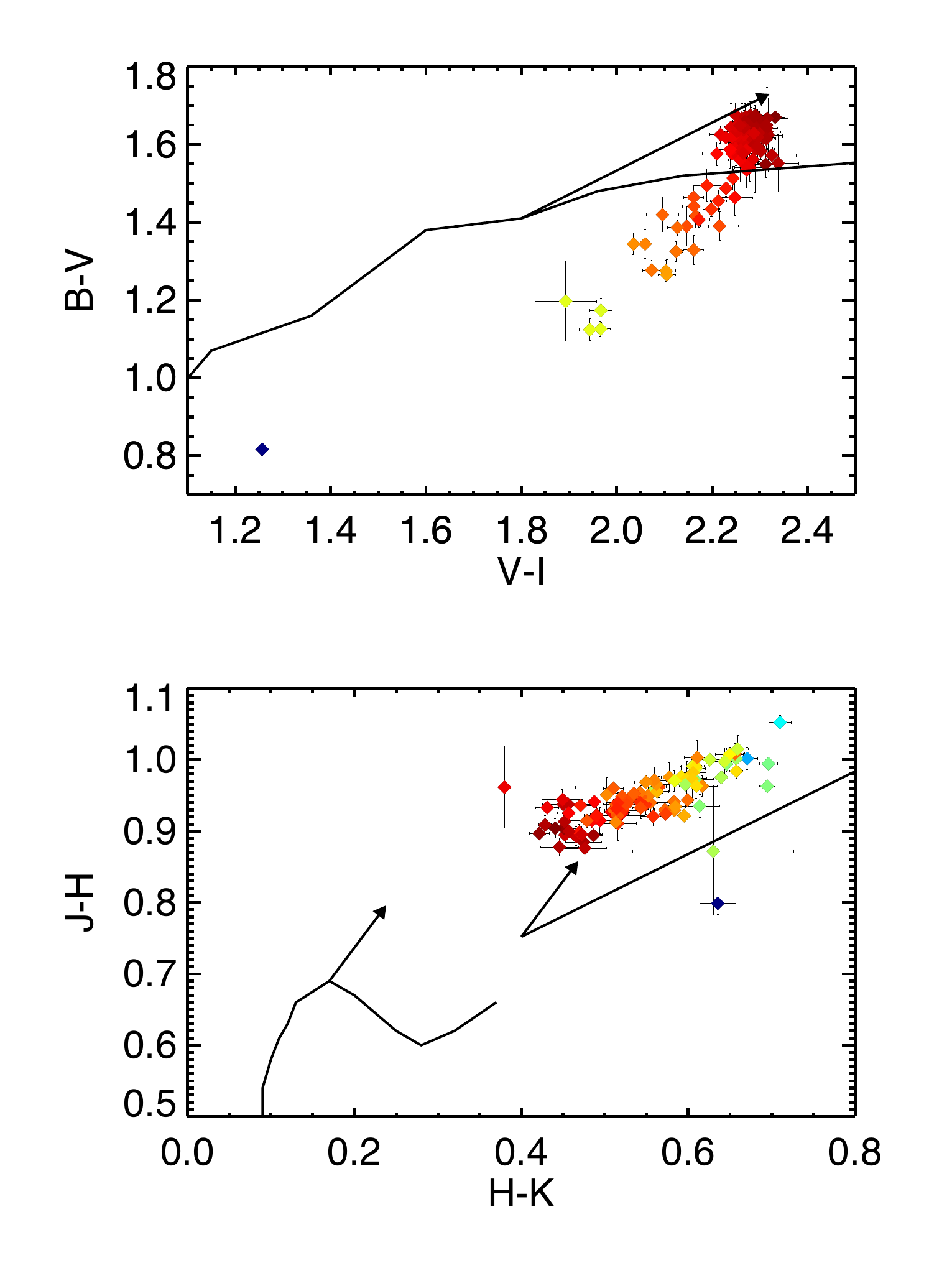}
\caption{Same as in Figure~\ref{smcc12}, for the 2013-2014 season.
\label{smcc13}}
\end{figure}

\begin{figure}[H]
\plotone{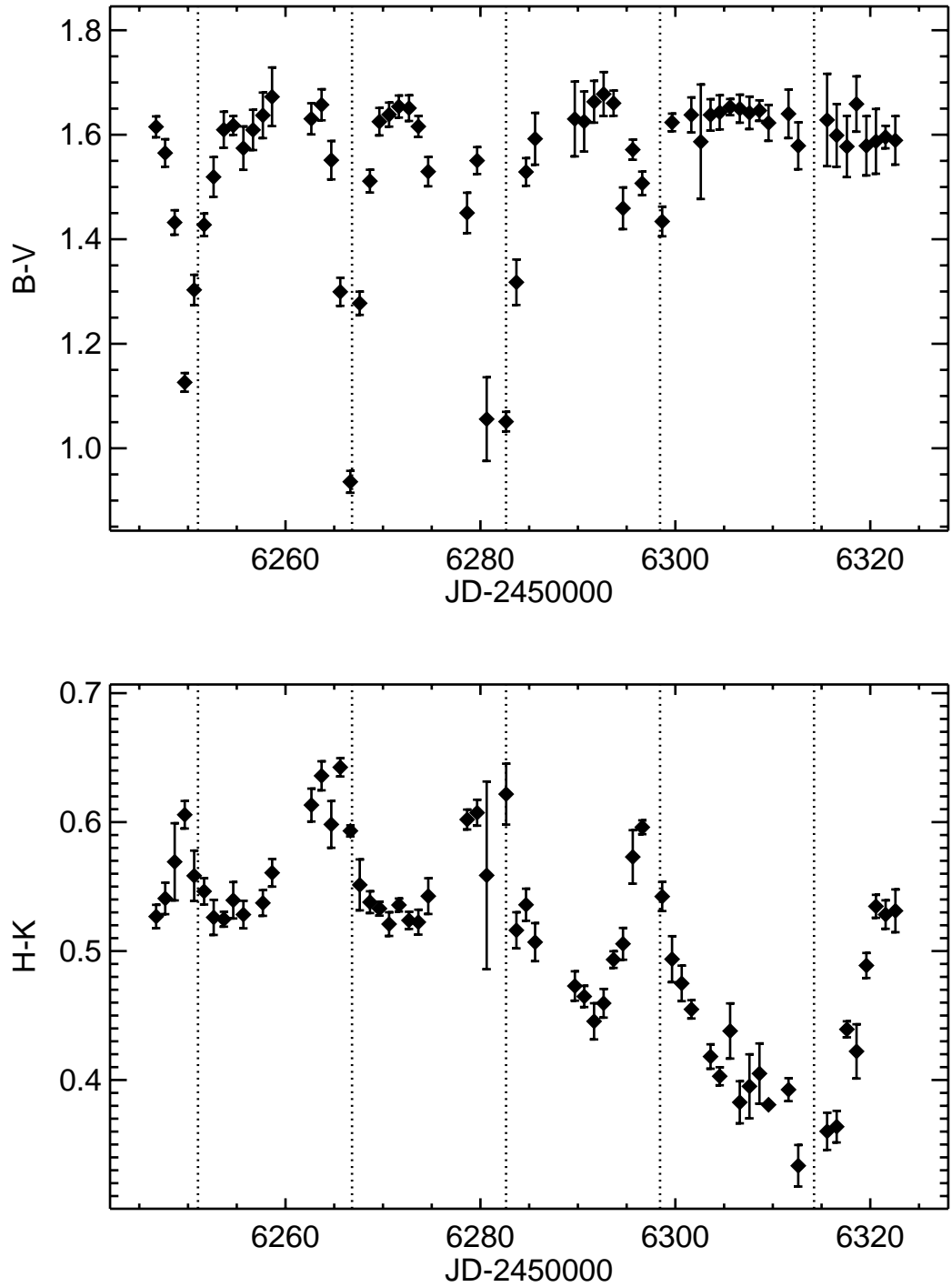}
\caption{SMARTS/ANDICAM {\it B-V} and {\it H-K} colors versus time for
the 2012-2013 season.
\label{smct12}}
\end{figure}

\begin{figure}[H]
\plotone{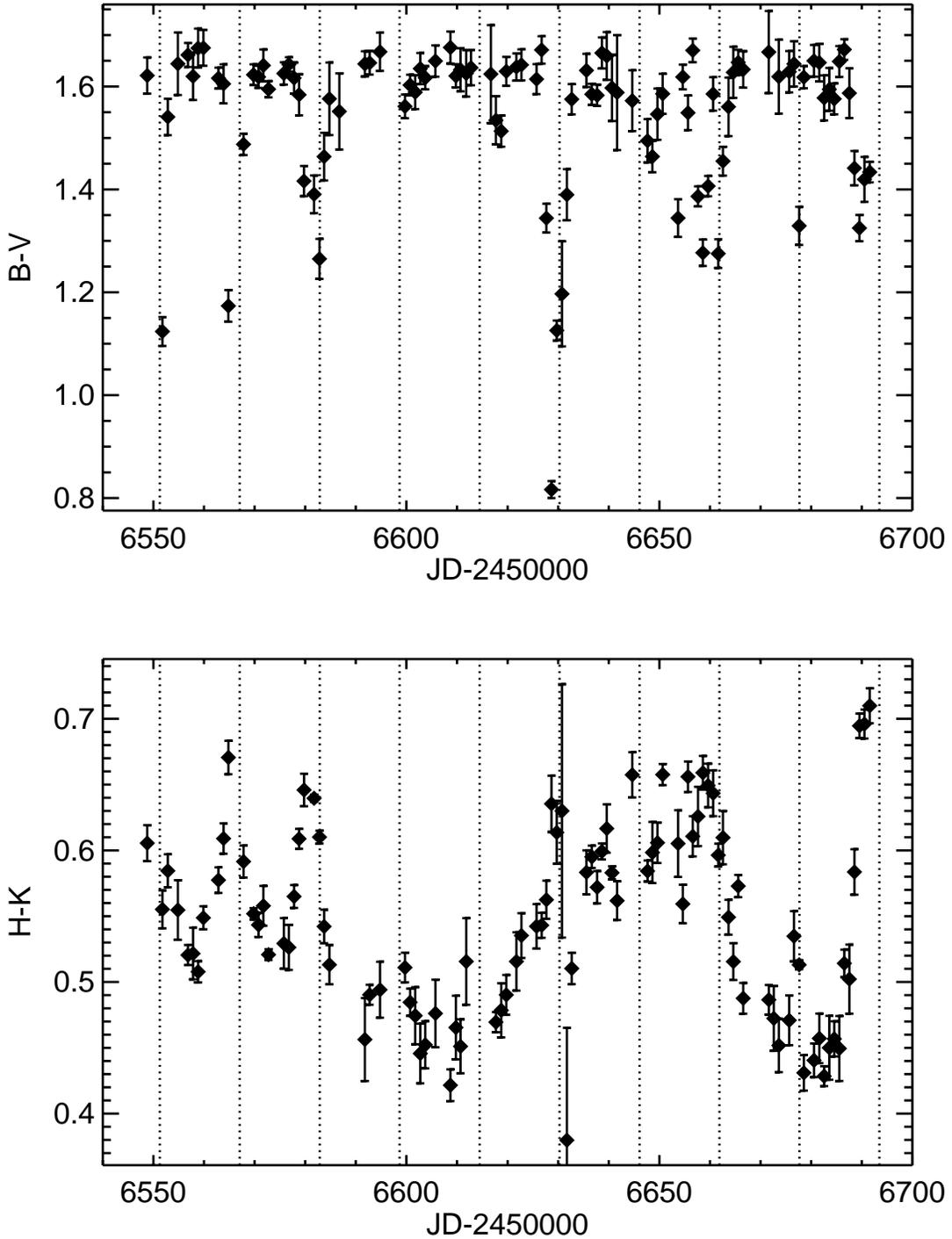}
\caption{Same as in Figure~\ref{smct12}, for the 2013-2014 season.
\label{smct13}}
\end{figure}

The pulse heights and, to a lesser extent, durations, are wavelength-dependent. 
Figures~\ref{smcc12} and~\ref{smcc13} show optical and NIR
color-color plots for both observing seasons.  In the optical, the flux level
is strongly correlated with $B-V$ color, with brighter epochs (i.e., the pulses)
being bluer. Plotted as a function of time, the pulse correlation can be seen
more clearly (Figs.~\ref{smct12},~\ref{smct13}).  The $B-V$ baseline between
the pulses is relatively flat, with the average color being consistent
with a reddened M0 photosphere with A$_V \gtrsim 1$; there is likely some
residual continuum excess at least at B and V bands, given that previous
measures of optical veiling rarely if ever decreased to zero
(e.g., Basri et al. 1997),
and the fact that small color variations do occur.

In the NIR, the brighter epochs correspond to {\it redder} colors.
In general, the NIR observations form a locus of points that are roughly
parallel to the CTTS locus (as defined by Meyer et al. 1997).  Dereddening
these points down to the locus (assuming that the separation is entirely due
to extinction, which may not be the case) yields a typical extinction value of
A$_V \sim 1.5$.  There are a few interesting outliers in the 2013-2014 season,
which we describe in more detail below.  The $H-K$ color time series
(lower panels of Figs.~\ref{smct12} and~\ref{smct13}) exhibit
peaks corresponding to each pulse event; most of these peaks are broader
in time, with an earlier start time in terms of binary phase (typically
at or near apastron phase), compared to the peaks seen in the optical.
They also typically reach a maximum earlier than the optical color peaks,
consistent with the behavior in the corresponding photometric bands
as described above.  In addition to the pulse events corresponding
to each orbital period,
the NIR photometry also exhibits a longer-term trend at H and K bands
with characteristic timescales of a few months.  The trend is most apparent
in the H-K color time series, exhibiting an amplitude of $\sim 0.15$
magnitudes.

\begin{figure}[H]
\includegraphics[scale=0.6]{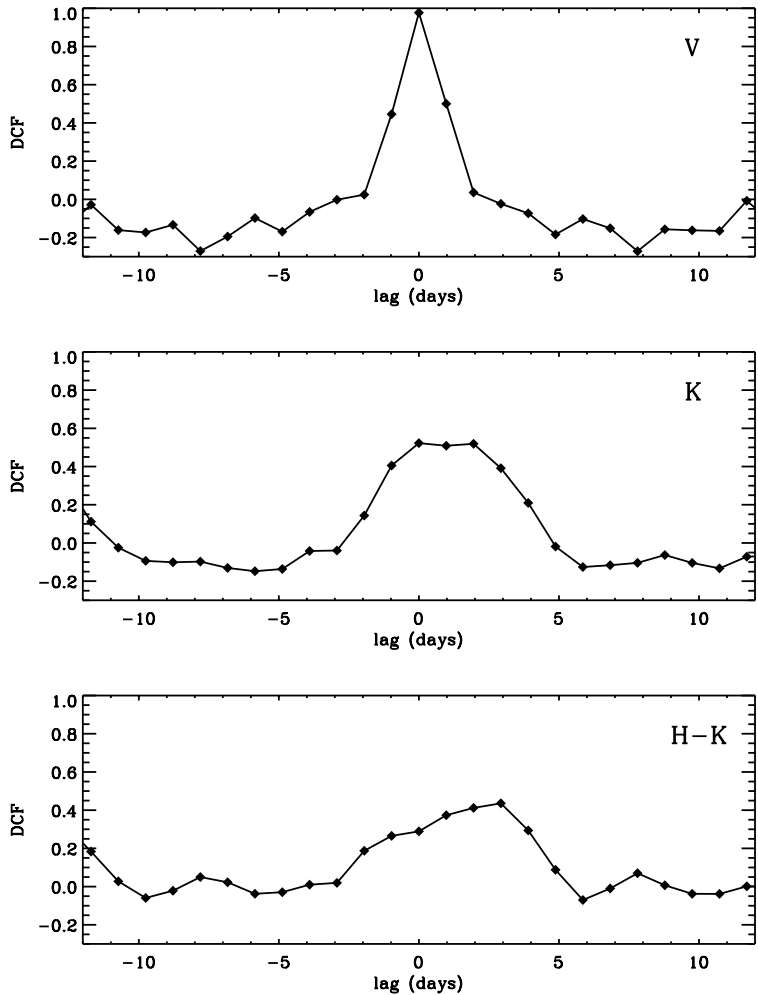}
\caption{Discrete correlation functions for the indicated bands compared
to the B band.  The photometry from both 2012-2013 and 2013-2014 seasons
were included.
\label{dcfs}}
\end{figure}

In order to quantify the apparent lags between bands, we computed
discrete correlation functions \citep[DCFs;][]{1988ApJ...333..646E} for the
combined 2012-2013 and 2013-2014 photometry.
This method is designed for unevenly-sampled data, as is the case here
(we also looked at cross-correlations of the light curves interpolated onto
a regular time spacing and obtained similar results).
Figure~\ref{dcfs} shows the DCFs of V and K bands and the H-K color
relative to the B band.  The V band is well-correlated with B and shows
no measurable lag.  K band is also correlated, though at a somewhat lower
level of significance.  Its DCF is clearly offset from zero lag, with maxima
at zero and 2-day lags and a "centroid" value at about a 1-day lag,
and is also broader than for V band.  This may be a further indication of
the longer apparent duration of the K band flux peaks on average compared
to the optical pulse events, or alternatively may be explained by two
separate events in the light curves, one with zero lag associated with
the optical events and one with lag $\sim 2$ days.
The H-K color DCF exhibits a weak peak at a lag of 3 days, however
the significance is not as robust given the smaller amplitude of
the H-K peaks.

\begin{figure}[H]
\includegraphics[scale=0.6]{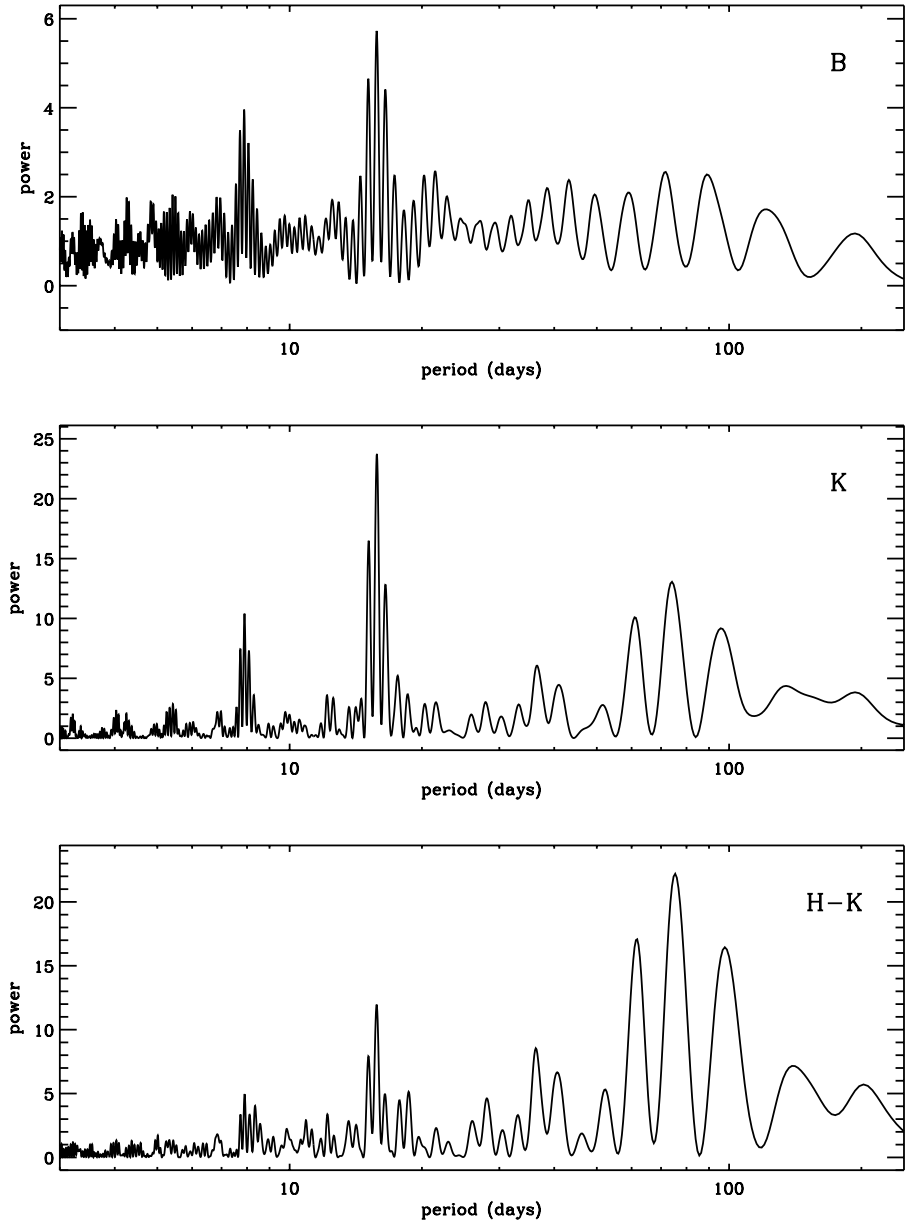}
\caption{Lomb normalized periodograms for the B, K, and {\it H-K} color
time series for the combined 2012-2013 and 2013-2014 seasons.  Note that
each of the major peaks are accompanied by several sidelobes, which are
artifacts caused by the large gap between the two seasons.
\label{pds}}
\end{figure}

Periodograms of the combined data from both seasons are shown
in Figure~\ref{pds}.  Results for the B band photometry show a
weak peak at 15.77 $\pm 0.41$ days (error estimated from the FWHM of
the periodogram peak), with false alarm probability (FAP) of 0.65,
indicating only a weak statistical correlation of the blue pulses with
the binary orbital period.  The K band periodogram also shows a peak
at 15.80 $\pm 0.42$ days, with FAP $\sim 2 \times 10^{-8}$,
a much stronger statistical correlation
that reflects the more steady nature of the NIR pulses;
there is also a second weaker but possibly significant peak at 74.1 days,
which corresponds to the long-term trend mentioned above.
The H-K color time series periodogram shows a strong peak at a similar
period of 75.4 $\pm 10.9$ days, with FAP $\sim 7 \times 10^{-8}$,
and a peak at 15.77 days
is still present but with lower significance.  Given that our data cover
only about three full periods of the 75-day feature,
confirmation of a persistent periodic feature
in the NIR light curves requires further long-term monitoring.
The 15.8-day periodicity exactly matches the binary orbital period
as measured from radial velocity observations.

There is also a higher-frequency low-amplitude variation seen during
several quiescent cycles in both seasons, with amplitude
decreasing to longer wavlengths.  Periodograms for those parts of the V-band
light curves (in the intervals JD 2456300-6325 and 2456585-6615) show
a peak at 3.0 $\pm 0.3$ days (with significance values of 0.05 and
0.09, respectively).  This is similar to the estimate of
the stellar rotation period of $\sim 3$ days from Basri et al. (1997)
based on measurements of $v$sin$i$, and is in agreement with the period
recently found from Kepler K2 observations
(3.017 days; K\'osp\'al et al. 2018).
This feature is likely the signature
of a rotating hot- or cold-spot on the surface of one or both of the stars,
a signature that is typically obscured by the larger-amplitude variations
likely driven by accretion.

\begin{figure}
\plotone{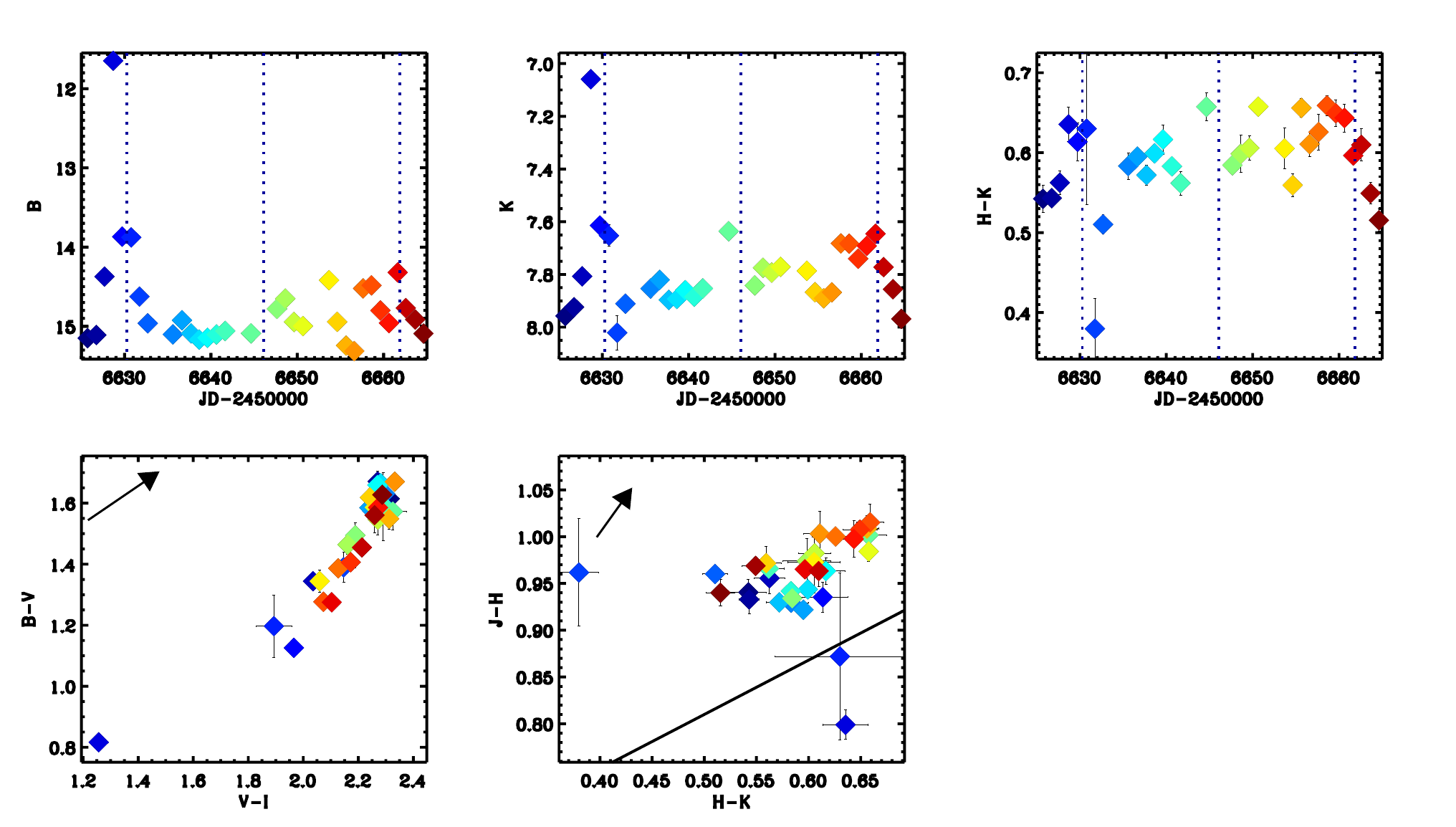}
\caption{A zoomed-in portion of the light curve from the 2013-2014 season,
along with various associated colors.  The symbols are color-coded according
to time to facilitate matching between panels.  Dashed lines indicate times of
periastron passages.  Arrows in the color-color plots indicate reddening
vectors for $A_V = 1$.  The solid line in the $H-K$ vs $J-H$ color-color plot
is the CTTS locus.
\label{2013zoom}}
\end{figure}

Several orbital cycles monitored in the 2013-2014 season presented particularly
noteworthy behavior, some of which contradicts the general photometric trends.
Figure~\ref{2013zoom} shows the multiband photometry and colors for this interval,
which spans about 2.5 cycles.  The most obvious feature is the extremely
large pulse near JD 2456630, as previously mentioned.  To our knowledge,
this is by far the largest amplitude optical brightening ever observed in DQ Tau,
with a maximum increase of about 3.4 magnitudes at $B$ and just under
1 magnitude at $K$.  The pulse peak is extremely blue in the optical,
consistent with emission from the accretion shock, which is so strong that even
the $J-H$ color is significantly bluer than at any other epoch (dropping below the
CTTS locus).  By contrast, the peak $H-K$
colors are similar to those of other pulse peaks.
Immediately after this strong pulse, the IR emission decreases significantly,
with the $H-K$ color exhibiting a particularly dramatic drop to one of the lowest
levels we see at any epoch, though it recovers quickly in just a few days.
The subsequent two orbital cycles show very different behavior.  No obvious
optical pulse appears before the next periastron passage, although there is a peak
in the NIR.  After that, the light curves become much more complicated;
there are {\it four} peaks in the optical separated by roughly four to five days,
and three peaks in the $H-K$ color curve (not coincident with the optical peaks).

\subsection{Spectroscopy}

\begin{figure}[H]
\includegraphics[scale=0.8]{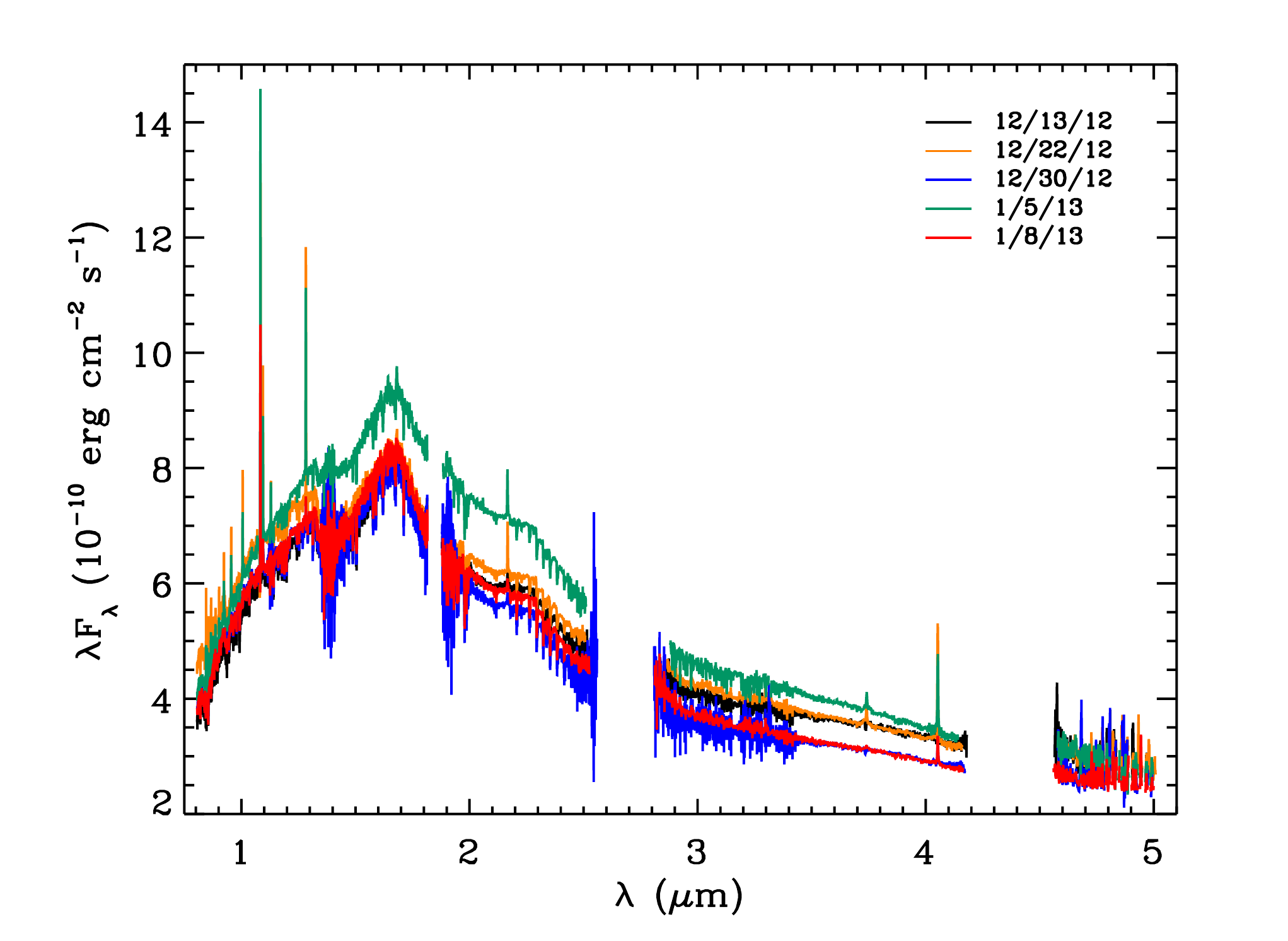}
\caption{IRTF/SpeX spectra obtained on the indicated nights.
Each spectrum has been scaled by the average relative offset
from the contemporaneous SMARTS photometry.
\label{spec}}
\end{figure}

The full set of spectra are shown in Figure~\ref{spec}.  The data quality
is uniformly high, with $S/N > 50$ over the entire wavelength range
except for the edges of some of the telluric absorption bands and parts
of the 4.6-5 $\mu$m order.  In most of the 1-2.5 $\mu$m range,
the $S/N > 300$.  Telluric absorption residuals are seen in some epochs
at the edges of the telluric windows and at 2.8-3.4 and 4.6-5 $\mu$m,
typically the result of an imperfect match of the telluric template air mass.
The spectra show absorption features typical of young early M-type stars,
as well as a wide range of emission lines whose strengths vary considerably
with epoch.  Our primary objective with these data is to derive spectra
of the dust excess emission in order to characterize its strength and shape.
This requires matching with a proper photospheric template and correcting
for extinction and veiling, a process which we describe in the next
two subsections.

\subsubsection{Spectral type and template matching}

The spectral type of DQ Tau as estimated from optical spectroscopy ranges
from about K5 \citep{1997AJ....114..781B} to M0.6 \citep{2014ApJ...786...97H}.
However, the SpeX data exhibit many features,
in particular broad molecular absorption bands such as TiO and H$_2$O,
that are more consistent with a later type.  This was also shown
by Bary14, who suggested that the discrepancy could be
explained by including the effects of large cool spots.  One difficulty
with deriving accurate spectral types in the infrared is that many
photospheric features are sensitive to surface gravity; the usual practice
of adopting main sequence dwarfs to calibrate absorption line strengths can
then have significant errors depending on which features are being compared
\citep[e.g.,][hereafter, McClure13]{2013ApJ...769...73M}.

\begin{figure}[H]
\includegraphics[scale=0.8]{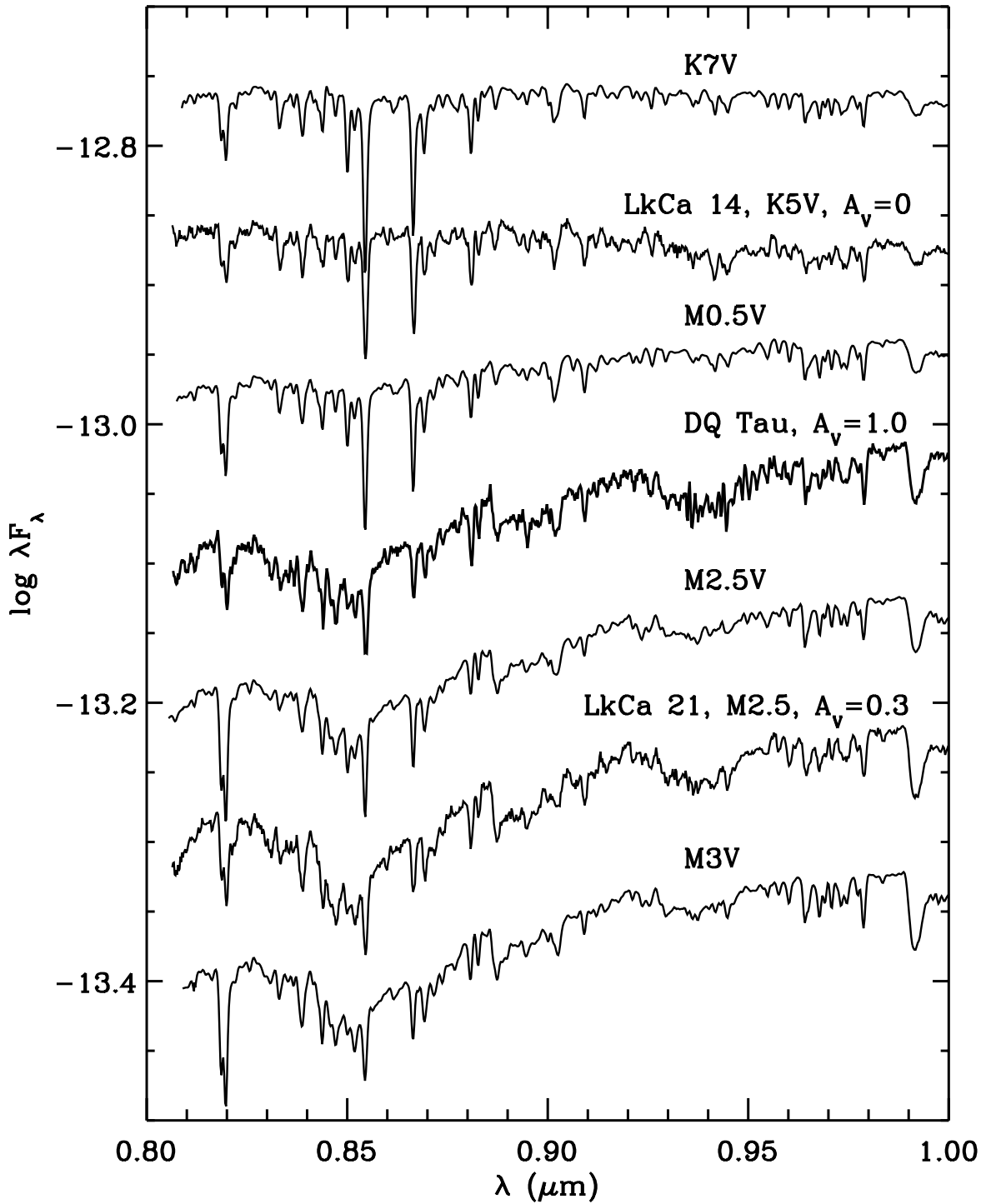}
\caption{Comparison of one of the DQ Tau SpeX spectra
(with negligible veiling in the displayed wavelength range)
to several template spectra with a range of spectral types.
Fluxes have been arbitrarily shifted to place each spectrum
in order of type.
Spectra labeled only with a spectral type are for dwarf stars
taken from the SpeX spectral library.
\label{templates}}
\end{figure}

\begin{figure}[H]
\plotone{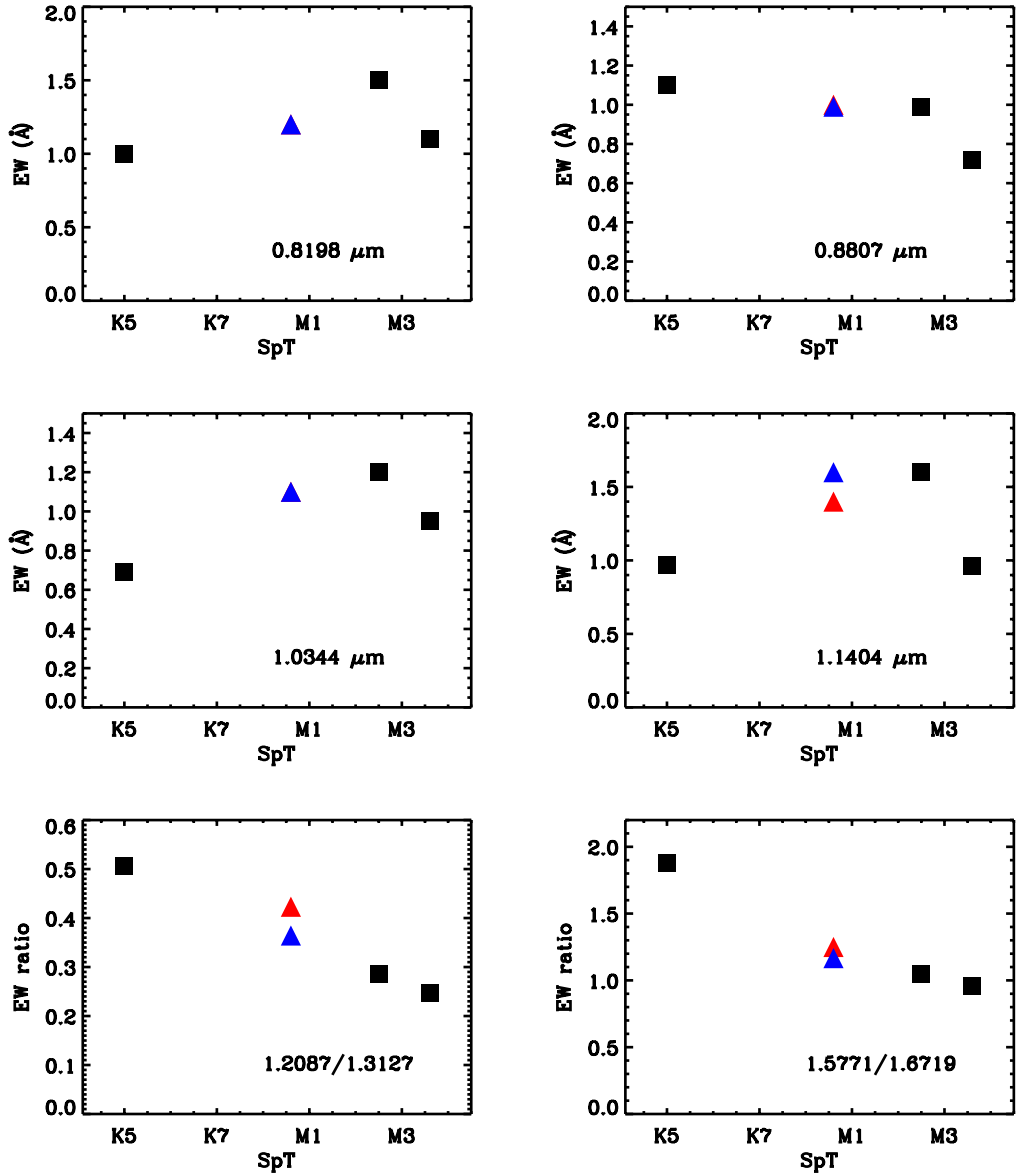}
\caption{Selected absorption line equivalent widths and line ratios as
a function of spectral type for two epochs of DQ Tau (red and blue triangles),
the WTTS templates (black squares),
and dwarf spectra from the IRTF spectral library (error bars).
The plotted spectral types for DQ Tau and the WTTSs were taken from
\citet{2014ApJ...786...97H}, based on optical spectra.  The formal measurement
uncertainties for the TTSs are equal to or less than the symbol sizes.
\label{abs_ews}}
\end{figure}

To try to mitigate surface gravity effects, we observed several weak line
T Tauri stars (WTTSs) with a range of spectral types to use as photospheric
templates (Table~\ref{templatetab}).
Figure~\ref{templates} compares the 0.8-1 $\mu$m range of one DQ Tau epoch
(which had a low level of accretion activity, and hence veiling) to two
of our WTTS templates, as well as several dwarf standards taken from
the SpeX spectral library \citep{2005ApJ...623.1115C, 2009ApJS..185..289R}.
The closest match by eye
is the WTTS LkCa 21, which has an optical spectral type of M2.5
\citep{2014ApJ...786...97H}.  For a more rigorous comparison, we measured
absorption line equivalent widths and line ratios, restricted to lines at
shorter wavelengths and/or closely spaced in wavelength in order to mitigate
the effects of veiling.  These specific indicators were found by
McClure13 to offer the most accurate measures of spectral type.
The resulting measurements for the WTTS templates
and two epochs of DQ Tau (in which the veiling was lowest) are shown in
Figure~\ref{abs_ews}.  We also measured the same features in the IRTF spectral
library dwarf spectra, degrading the resolution in order
to match our observations taken with a wider slit.  In most cases, DQ Tau
and the WTTSs follow the trend with spectral type indicated by the dwarf
measurements.  The M3.6 WTTS LkCa 1 has systematically weaker absorption
for all but one of the four lines plotted, although the two line ratios
are consistent with its optical spectral type; this may be a surface gravity
effect (note the luminosity derived by Herczeg \& Hillenbrand is fairly
high for its spectral type, suggesting a larger radius and lower surface
gravity).  Overall, the measurements for DQ Tau suggest a spectral type
in the range M0-M1, in excellent agreement with the most recent
optically-derived type of M0.6 from \citet{2014ApJ...786...97H}.
Again, of the three WTTS templates the closest match is LkCa 21.
Given the reasonable similarity in absorption line equivalent widths as well as
the excellent match to the broader TiO and H$_2$O absorption bands,
we adopted LkCa 21 as the photospheric template in all further analyses
of the dust excess emission in DQ Tau.

\begin{deluxetable}{lccc}
\tablewidth{0pt}
\tablecaption{WTTS templates}
\tablehead{
\colhead{Object} &
\colhead{spectral type} &
\colhead{$A_V$} &
\colhead{log L ($L_{\odot}$)}}
\startdata
LkCa 14 & K5.0 & 0.0 & -0.15\\
LkCa 21 & M2.5 & 0.3 & -0.37\\
LkCa 1 & M3.6 & 0.45 & -0.29
\enddata
\tablecomments{All measurements from \citet{2014ApJ...786...97H}.}
\label{templatetab}
\end{deluxetable}

\subsubsection{Derivation of excess spectra}

The photospheric absorption lines in our spectra of DQ Tau are subject to
continuum veiling, particularly at wavelengths $\gtrsim 1.5 \mu$m.
This excess emission originates largely from hot dust in
the innermost regions of the system, as indicated by our NIR photometry.
By characterizing the excess as a function of wavelength at multiple epochs,
we can infer general properties such as temperature, location, and size of
the emitting region, and how these vary with time.

To measure the veiling, we compared depths of selected absorption lines in
each spectrum of DQ Tau with those of the template, LkCa 21.  There are
several methods that have been devised originally for optical measurements
\citep[e.g.][]{1989ApJS...70..899H} that are equally
applicable to the NIR (McClure13).  We adopted the equivalent width method,
whereby the veiling is measured for individual absorption lines using
the ratio of equivalent widths between the object and template spectra:
$EW_{temp}/EW_{obj} = 1 + r_\lambda$.  As noted by McClure13, this technique
has the advantage of being able to ignore lines that may be sensitive to
differences in surface gravity or cool spot surface coverage between the
object and stellar template.  The measured line equivalent widths for all
DQ Tau epochs and two WTTSs are given in Table~\ref{abs_ew}.
We used only atomic lines at wavelengths shorter than $\sim 2.5 \mu$m;
molecular features such as the CO overtone lines around 2.3 $\mu$m
are more sensitive to gravity and spot effects,
and any lines at longer wavelengths are filled in by the veiling. 

\begin{deluxetable}{lccccccc}
\tabletypesize{\small}
\tablewidth{0pt}
\tablecaption{Absorption line equivalent widths}
\tablehead{
\colhead{line} &
\colhead{} &
\colhead{} &
\colhead{DQ Tau} &
\colhead{} &
\colhead{} &
\colhead{LkCa 21} &
\colhead{LkCa 14} \\
\colhead{} & 
\colhead{12/13/12} &
\colhead{12/22/12} &
\colhead{12/30/12} &
\colhead{1/5/13} &
\colhead{1/8/13} &
\colhead{} &
\colhead{}}
\startdata
0.81980 & 1.20 & 1.10 & 1.20 & 0.97 & 1.20 & 1.50 & 1.00\\
0.83840 & 1.20 & 0.92 & 1.40 & 0.77 & 1.30 & 1.50 & 1.00\\
0.88070 & 1.00 & 0.65 & 0.99 & 0.67 & 0.89 & 0.99 & 1.10\\
0.90900 & 0.61 & 0.40 & 0.60 & 0.40 & 0.57 & 0.60 & 0.78\\
0.97880 & 0.95 & 0.70 & 0.92 & 0.72 & 0.81 & 0.94 & 0.76\\
1.03440 & 1.10 & 0.95 & 1.10 & 0.85 & 1.00 & 1.20 & 0.69\\
1.14040 & 1.40 & 1.10 & 1.60 & 1.00 & 1.40 & 1.60 & 0.97\\
1.16890 & 0.81 & 0.71 & 0.81 & 0.61 & 0.79 & 0.79 & 0.56\\
1.18300 & 0.99 & 0.80 & 0.99 & 0.71 & 1.00 & 0.99 & 1.60\\
1.2087 & 0.55 & 0.42 & 0.51 & 0.36 & 0.55 & 0.40 & 0.71\\
1.25250 & 0.80 & 0.68 & 0.85 & 0.58 & 0.88 & 0.85 & 0.50\\
1.31270 & 1.30 & 1.20 & 1.40 & 1.10 & 1.30 & 1.40 & 1.40\\
1.31500 & 1.00 & 0.90 & 1.10 & 0.84 & 0.97 & 1.10 & 1.10\\
1.48818 & 1.80 & 1.50 & 1.90 & 1.30 & 1.80 & 1.90 & 2.70\\
1.50270 & 2.10 & 1.80 & 2.20 & 1.70 & 2.00 & 2.30 & 3.00\\
1.57710 & 2.00 & 1.90 & 2.10 & 1.70 & 2.00 & 2.30 & 3.20\\
1.58900 & 2.00 & 1.80 & 2.30 & 1.30 & 2.20 & 2.20 & 3.40\\
1.61550 & 0.94 & 0.82 & 0.99 & 0.81 & 0.93 & 1.10 & 1.60\\
1.62590 & 1.00 & 0.93 & 1.10 & 0.87 & 1.00 & 1.40 & 1.30\\
1.6719 & 1.60 & 1.80 & 1.80 & 1.30 & 1.60 & 2.20 & 1.70\\
1.67550 & 1.90 & 2.10 & 2.10 & 1.70 & 2.10 & 2.50 & 2.30\\
1.71130 & 2.60 & 2.60 & 2.80 & 2.20 & 2.80 & 3.50 & 3.20\\
2.11650 & 0.82 & 0.89 & 1.10 & 0.76 & 1.10 & 1.60 & 1.30\\
2.26000 & 1.90 & 1.50 & 2.10 & 1.40 & 1.90 & 3.20 & 2.70
\enddata
\tablecomments{First column is line wavelength in units of microns.
Equivalent widths are in units of {\AA}.}
\label{abs_ew}
\end{deluxetable}

In order to derive the excess emission spectrum at each epoch, the observed
spectrum must be correctly matched with that of the template.
We followed a procedure similar to \citet{1998ApJ...492..323G},
originally applied to UV/optical data and subsequently adapted to
NIR data by \citet{2011ApJ...730...73F} and McClure13.
This method allows a simultaneous derivation of the reddening and normalization
factor given the measured veiling.  In short, the ratio of the continum fluxes
of the object ($F_{obj}$) and the template ($F_{temp}$), modified by the
object veiling, is related to the difference in extinction between the two by
\begin{equation}
  2.5 \log
  \left[
   \frac{F_{\lambda,temp}}{F_{\lambda,obj}} (1+r_{\lambda})
  \right]
  = 2.5 \log \, C \, \frac{A_{\lambda}}{A_V}
  \left( A_{V,obj} - A_{V,temp} \right) ,
\end{equation}
where $C$ is the normalization factor between object and template,
and $A_{\lambda}/A_V$ is the extinction law.  By measuring the flux and
veiling at a number of lines across the spectrum, the left-hand side of
equation 1 can be evaluated. A linear fit between those values and
an assumed extinction law 
then yields the difference in extinction $A_{V,obj}-A_{V,temp}$
(from the slope of the fit), and the normalization constant $C$
(from the y-intercept).  We adopted the extinction law from \citet{1990ARA&A..28...37M}
for $\lambda < 3 \mu$m, and \citet{2007ApJ...663.1069F} for longer wavelengths.
Once the extinction and normalization factor have been determined, the object
spectrum can be dereddened, and the template can be scaled and subtracted
to remove the photospheric component.

\begin{figure}[H]
\epsscale{0.8}
\plotone{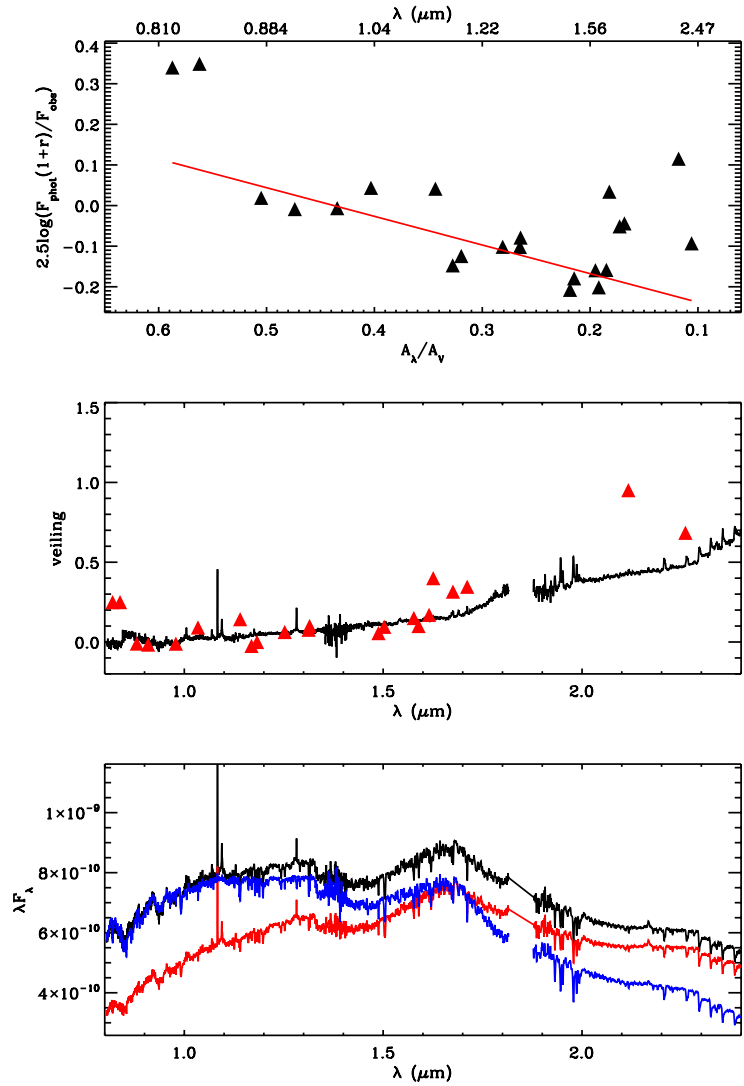}
\caption{Derived quantities and spectra from the 12/13/2012 observation of
DQ Tau.  (Top) Evaluated expression from equation 1, as a function of
the extinction law, at each of the absorption lines for which the veiling
was estimated (black triangles). A linear fit to the optimized subset of
points is shown in red.  (Middle) The derived veiling spectrum (black),
with the measured values for individual absorption lines overplotted
(red triangles).  (Bottom) Comparison of the normalized dereddened DQ Tau
spectrum (black), dereddened LkCa 21 template spectrum (blue),
and original unscaled DQ Tau spectrum (red).
\label{veiling1}}
\end{figure}

\begin{figure}[H]
\plotone{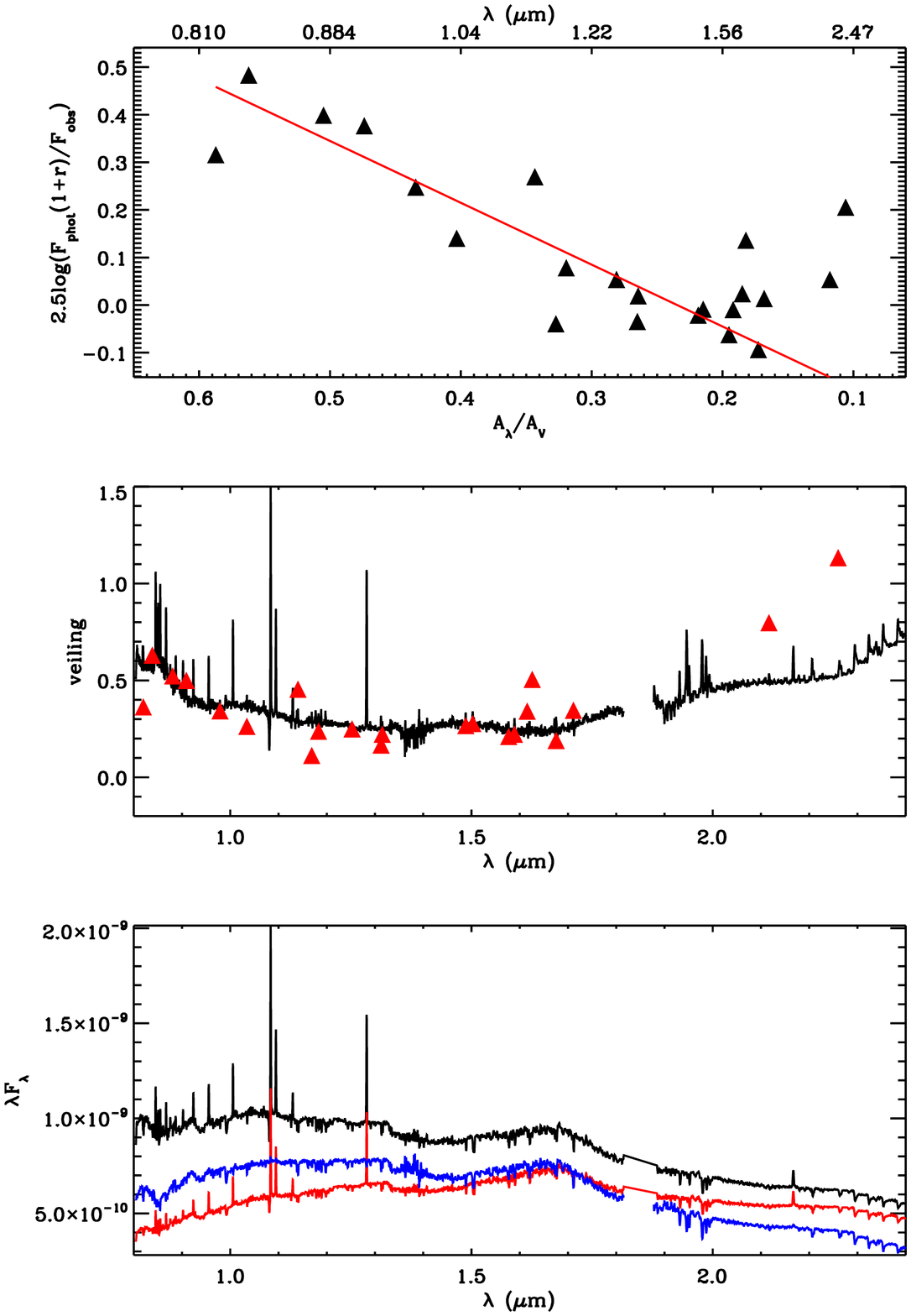}
\caption{Same as in Fig.~\ref{veiling1}, for the 12/22/2012 observation.
\label{veiling2}}
\end{figure}

\begin{figure}[H]
\plotone{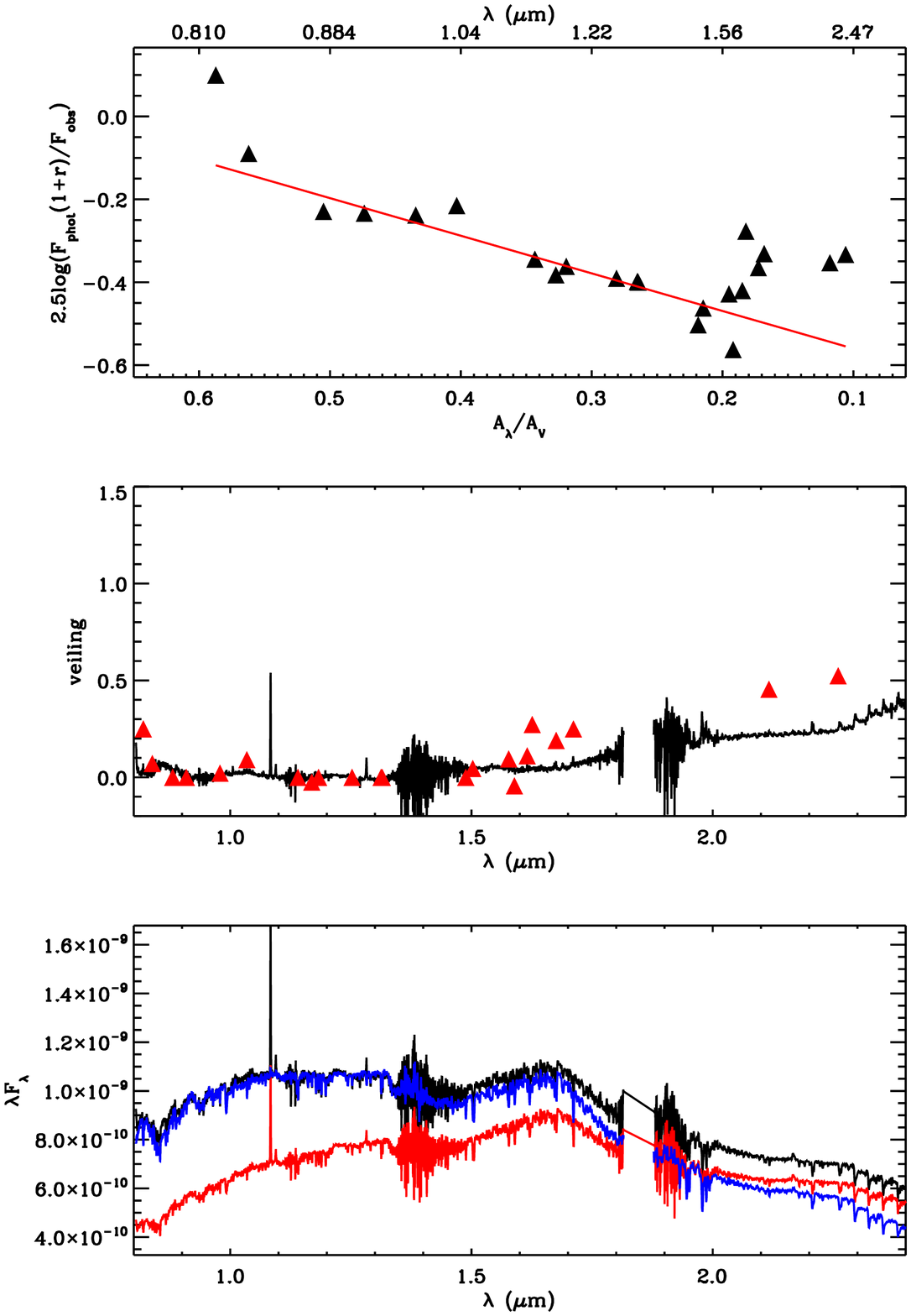}
\caption{Same as in Fig.~\ref{veiling1}, for the 12/30/2012 observation.
\label{veiling3}}
\end{figure}

\begin{figure}[H]
\plotone{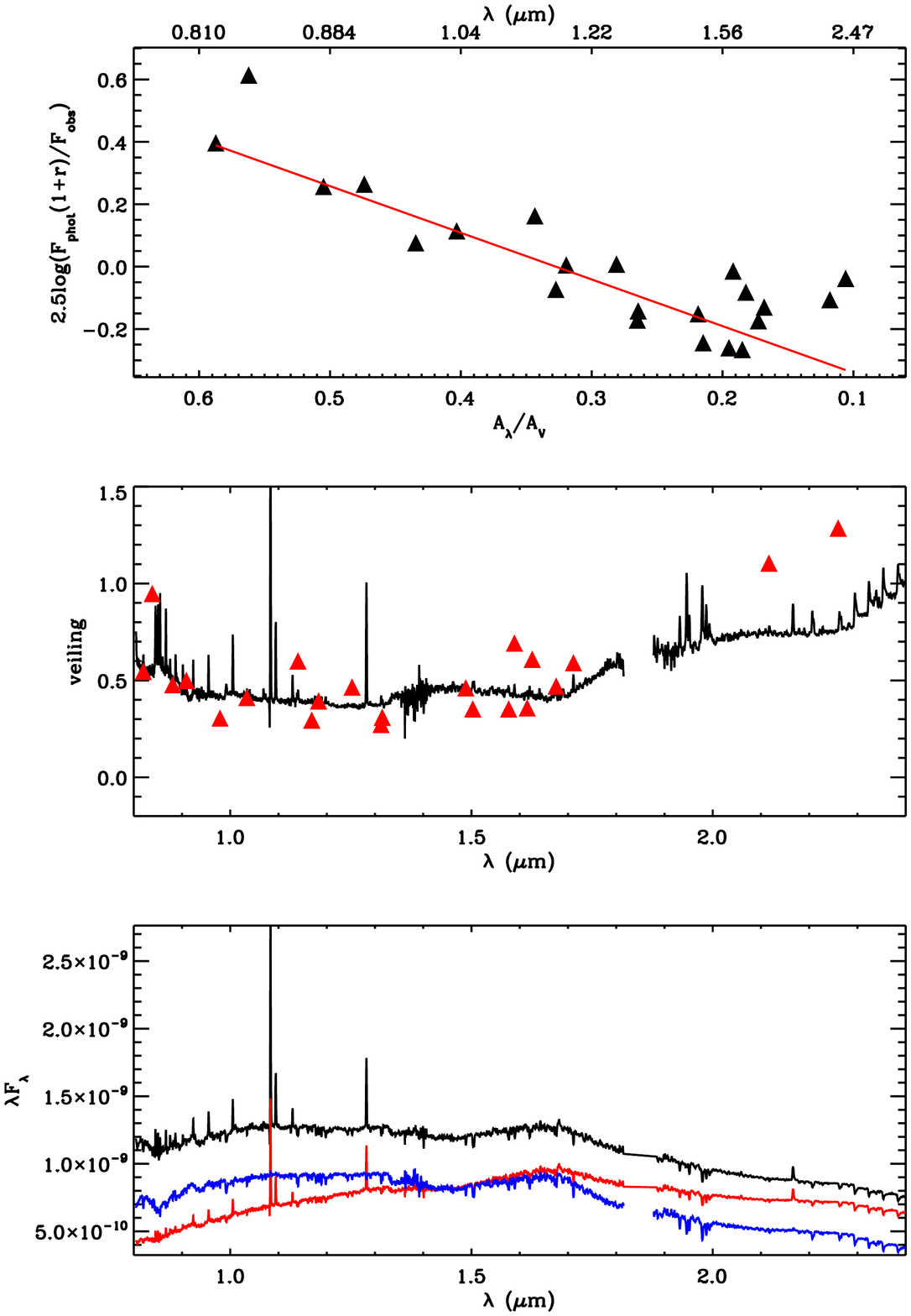}
\caption{Same as in Fig.~\ref{veiling1}, for the 1/5/2013 observation.
\label{veiling4}}
\end{figure}

\begin{figure}[H]
\plotone{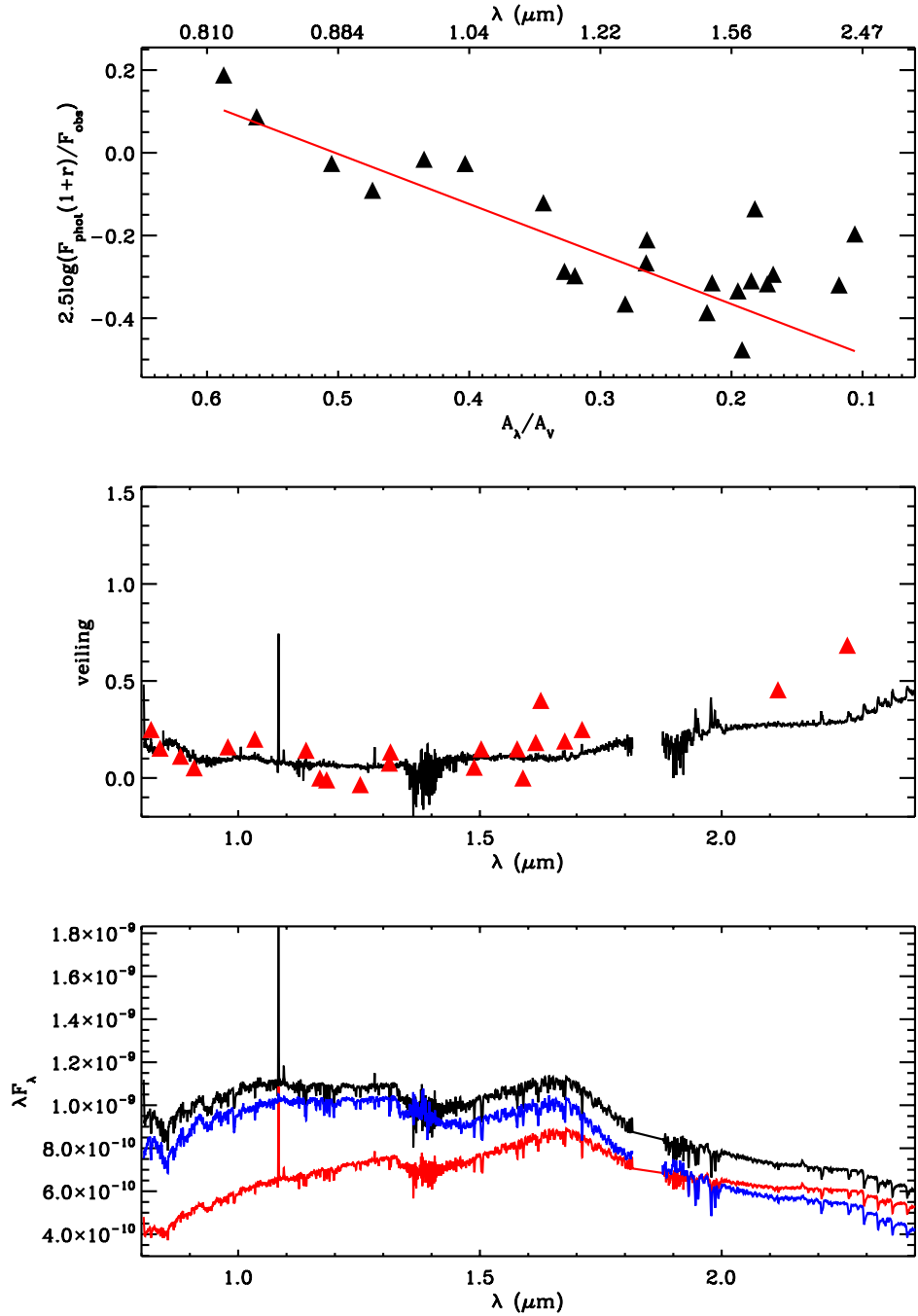}
\caption{Same as in Fig.~\ref{veiling1}, for the 1/8/2013 observation.
\label{veiling5}}
\epsscale{1}
\end{figure}

The results of these calculations for each observation epoch are shown in
Figures~\ref{veiling1}-~\ref{veiling5}.  Each upper panel plots
the evaluation of equation 1 for each of the absorption lines
indicated in Table~\ref{abs_ew}, along with a linear fit to the data points.
Some of the lines selected for this analysis are consistently poor matches to
the template.  In particular, the lines at 0.8198 and 0.8384 $\mu$m, and most
of the lines at wavelengths $> 1.6 \mu$m, are significantly stronger in
the template spectrum than in any of the DQ Tau spectra, and thus give
systematically larger veiling values.  We believe this is due to a mismatch in
the gravity and/or the fraction of the stellar surface covered by cool spots
between object and template, and elected to ignore these lines
when performing the linear fits.  The middle panels show the measured veiling
values for each line, along with a veiling spectrum calculated by dividing
the excess spectrum by the template.  There is good agreement between
the individual veiling values and the calculated spectrum for all except
the aforementioned discrepant lines, which indicates self-consistency between
the derived extinction and normalization constant.
Finally, the lower panels show the resulting dereddened and scaled
object spectrum compared to the original observed spectrum and the template.

The resulting extinction estimates for each epoch are given in Table~\ref{av}.
The values differ from each other, with a $>4 \sigma$ maximum deviation.
Moreover, it appears that the extinction is systematically larger closer
to periastron passages (see Discussion).
The extinction can be independently estimated from the maximum observed
$V-I$ color (in between pulse events), assuming no excess and an M0
photosphere; in that case, we derive $A_V \sim 0.9$, which is identical
within the uncertainties to the minimum spectroscopically-derived value.
The uncertainties quoted in Table~\ref{av} reflect the formal errors
on the fit, and do not take into account
possible systematic errors from any mismatch between the object
and template spectra.  A detailed comparison using a larger grid of templates
is needed to better constrain systematics, but we currently lack the sample
to do this.  Nevertheless, any such systematics should affect all epochs
equally, and we believe that the apparent extinction variations are likely real.

\begin{deluxetable}{lcc}
\tabletypesize{\small}
\tablewidth{0pt}
\tablecaption{\label{av}Extinction from fitted spectra}
\tablehead{
\colhead{date} &
\colhead{phase} &
\colhead{$A_V$}}
\startdata
12/13/12 & 0.51 & 1.01 $\pm 0.13$\\
12/22/12 & 1.08 & 1.60 $\pm 0.19$\\
12/30/12 & 0.59 & 1.21 $\pm 0.11$\\
1/5/13 & 0.97 & 1.80 $\pm 0.22$\\
1/8/13 & 1.16 & 1.51 $\pm 0.19$
\enddata
\end{deluxetable}

\subsubsection{Characterizing the excess}

\begin{figure}[H]
\includegraphics[scale=0.7]{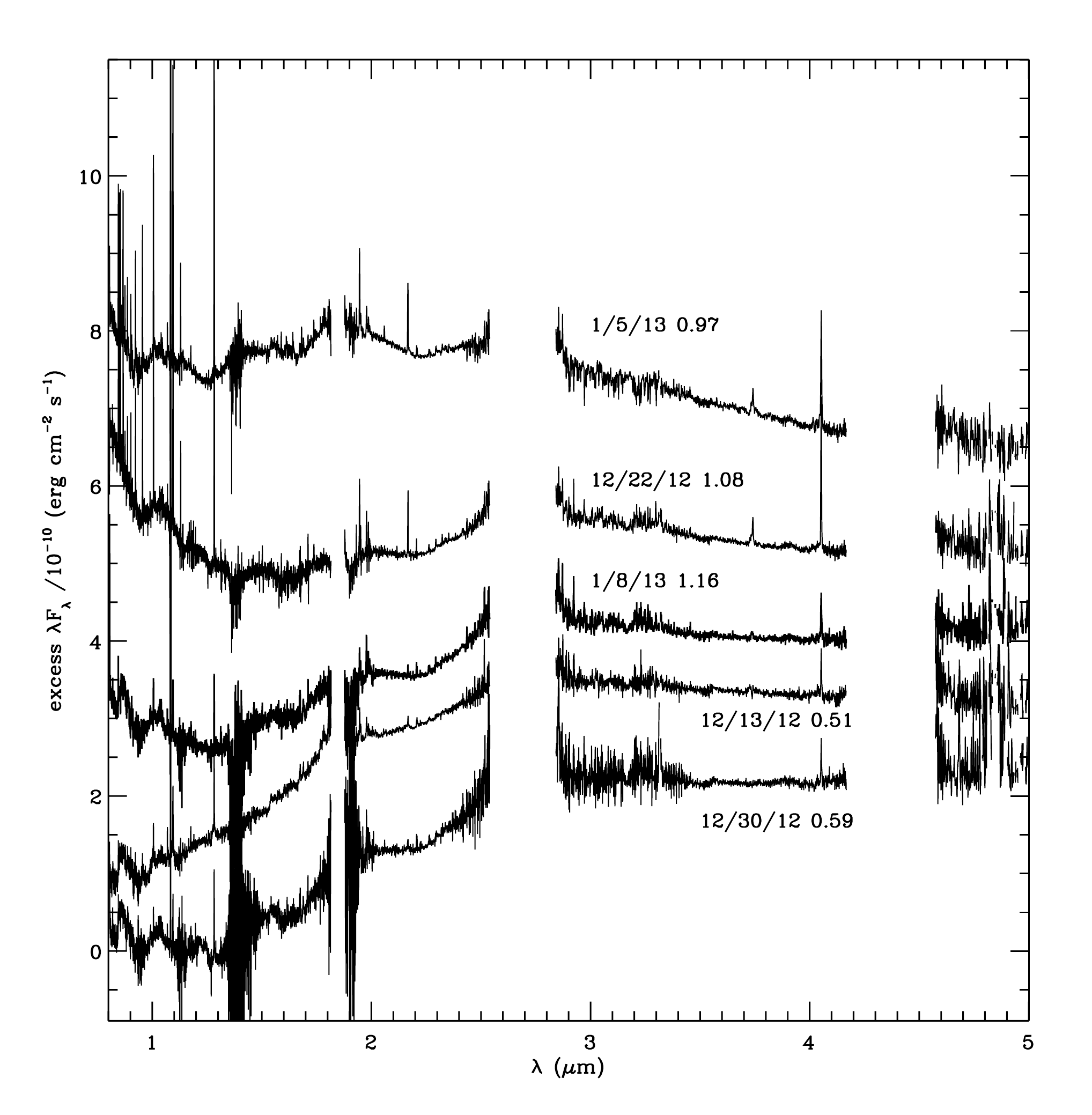}
\caption{The DQ Tau excess spectra derived from the five SpeX observations.
The observation date and corresponding binary orbital phase (where one 
is the time of periastron passage) are labeled in each case.
For clarity, the spectra have been offset along the y-axis
by the following amounts (from top to bottom): 4, 3, 2, 1, 0.
\label{excesses}}
\end{figure}

The excess spectra resulting from subtraction of the template from each
dereddened object spectrum are shown in Figure~\ref{excesses}.  There is
a clear change in both the shape and strength among the different epochs,
apparently correlated with the binary orbital phase.
The two observations taken closest to periastron passages show the largest
excess with the bluest colors, while the reverse is true for those taken
farthest in time from a periastron passage.  There is a similar correlation
with the emission line fluxes
(see next subsection).  The most dramatic
changes occur in the 0.8-1.5 $\mu$m range, which exhibits nearly zero
continuum excess near apastron phase but increases to a nearly flat excess
spectrum near periastron.  Other broad spectral features are also apparent
in some or all epochs, such as the ``bumps" centered roughly at 0.85, 1.05,
1.8, and 2.6 $\mu$m; these are coincident with molecular features such as
TiO and H$_2$O, and are most likely indicative of slight spectral mismatches
between object and template or, particularly in the latter case, artifacts
introduced by imperfect telluric correction.

In order to derive a rudimentary characterization of the excess shape,
we fit a set of blackbody components to each observation. The optical-NIR
continuum excess in most CTTSs has several components: ``hot" emission
most likely originating from the accretion shock on the stellar surface,
``warm" emission produced by dust at or near the the inner disk rim where
temperatures reach the sublimation point,
and ``cool" emission produced by the region of the disk behind the inner rim.
Detailed studies have indicated that the picture is somewhat more complicated,
with intermediate-temperature components of uncertain origin
\citep[e.g.][]{2011ApJ...730...73F}.  For the purposes of this study, where we are
taking a first look at time variability in the excess, we restrict ourselves
to blackbody models of these three basic components, following McClure13.
Each component is defined by a characteristic blackbody temperature
and solid angle.  We fix the hot component temperature
at 7000 K, which is characteristic of accretion shock emission
(the exact value is not crucial here since we are only seeing the Rayleigh-Jeans
tail).  We also fix the cool component to a value of 600 K, which is roughly
the equilibrium temperature for dust at the predicted location of the inner
edge of the circumbinary disk around DQ Tau ($\sim 2.5a = 0.35$ AU).
For the warm component, we allow the temperature to vary between 1000-1800 K
(which spans the range of typical characteristic maximum dust temperatures
in TTSs) in increments of 50 K.  The solid angles (relative to the stellar
value assuming $R=1.5$ R$_\odot$ and $d=196$ pc)
are varied from 0 to 0.1 for the hot component,
10 to 50 for the warm component, and 50 to 300 in the cool component.

\begin{figure}[H]
\includegraphics[scale=0.9]{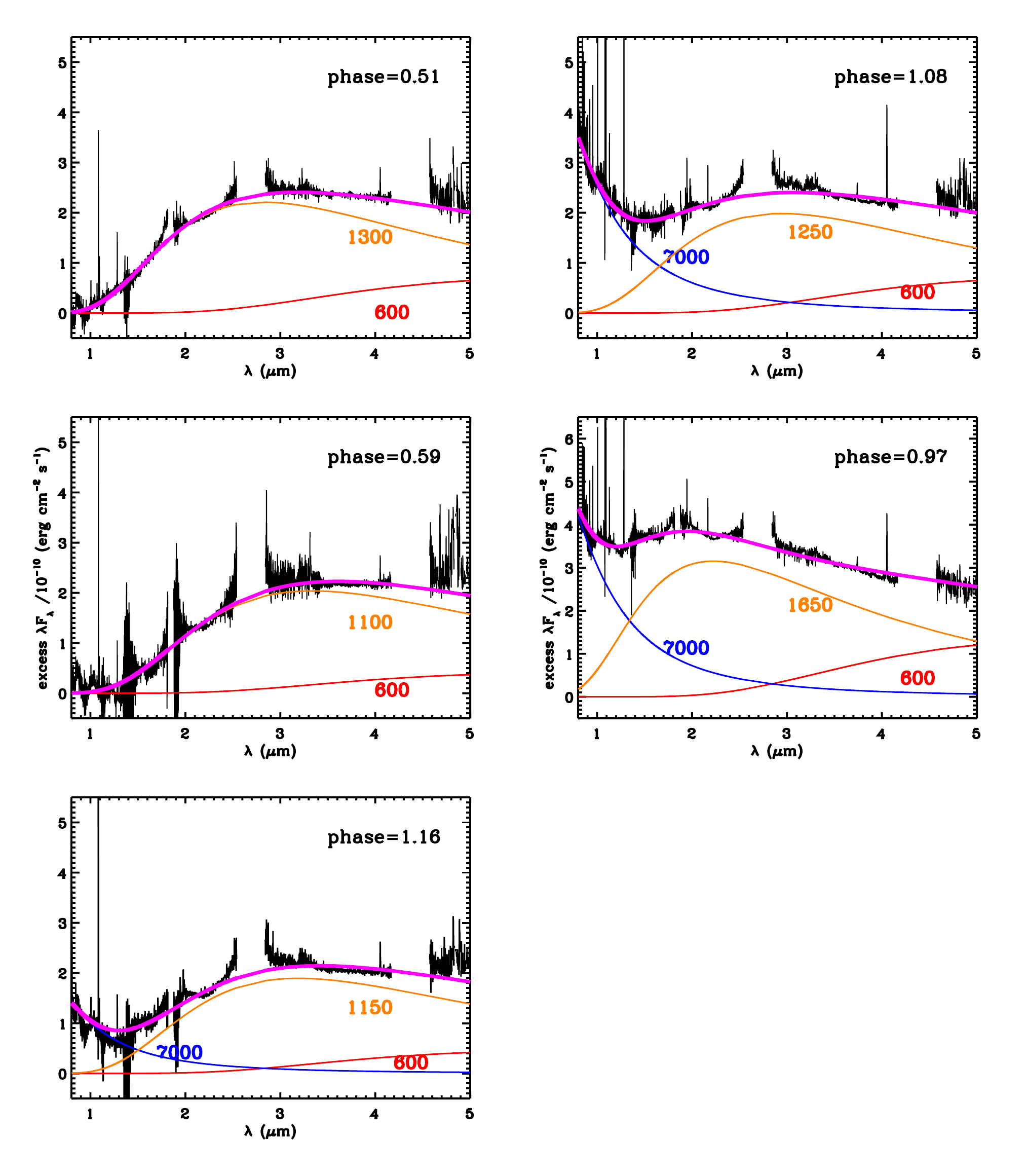}
\caption{DQ Tau excess spectra shown in black, in order of observation
time (left to right, top to bottom), with the binary orbital phase indicated.
The top two panels and bottom three panels show results from
the same binary orbit.  Colored lines represent the best-fit blackbody models,
with the three separate components and their characteristic temperatures
shown in blue, orange, and red, and the combined model shown in magenta.
\label{excesses_withfits}}
\end{figure}

For each set of parameters, the models for all three components are combined
and then compared to the observed excess spectra, with the best fit determined
via chi-squared minimization.  The resulting best fits for each observed
spectrum are shown in Figure~\ref{excesses_withfits}, with the associated
parameters listed in Table~\ref{model_param}.  Given the systematic
uncertainties in the template matching, these values should be taken as
indicative only; however, they provide a useful first look at the gross 
properties of the material in the inner regions of this system.
The best-fit hot component
solid angles range from zero at several epochs, where no discernible emission
was detected, to 0.06, which is typical for weakly- to moderately-accreting
TTSs.  The cool component solid angles range from 90 to 260,
roughly as expected for a 
puffed inner disk rim with a temperature of 600 K.
If our assumption of a constant temperature is a reasonable approximation,
the variation in solid angle may be indicative of a variable scale height
of the inner edge of the putative circumbinary disk.
There is a weak correlation with the warm component,
but the origin of such a variation is unclear; it could simply be
an artifact of our overly-simplistic assumptions of the dust temperatures.
However, it should be noted that this cooler dust emission is not
as well-constrained as the other components
since it contributes an appreciable fraction of the total emission only at
the long-wavelength end of our spectra. Observations in the 5-10 $\mu$m range
are required to test the significance of any variations from this region.

The warm component presents the most obvious variations.
The best-fit temperatures range from 1100 K, lower than typical dust
sublimation temperatures, to 1650 K, which likely corresponds to dust
at the sublimation front.  The hottest value coincides
with a NIR photometric peak several days before a periastron passage,
while the coolest value coincides with an apastron passage.
The temperatures are also cooler at and just after periastron passage.
This suggests a sequence of increasingly warming dust as the binary
moves closer together in its orbit, with the hottest dust being removed
during or just after the accretion pulse.

\begin{deluxetable}{lccccccc}
\tabletypesize{\small}
\tablewidth{0pt}
\tablecaption{Best-fit parameters for blackbody fits}
\tablehead{
\colhead{date} &
\colhead{phase} &
\colhead{hot T$_{eff}$} & 
\colhead{hot $\Omega$} &
\colhead{warm T$_{eff}$} & 
\colhead{warm $\Omega$} &
\colhead{cool T$_{eff}$} & 
\colhead{cool $\Omega$}}
\startdata
12/13/12 & 0.51 & 7000 & 0 & 1300 & 20 & 600 & 140\\
12/22/12 & 1.08 & 7000 & 0.05 & 1250 & 21 & 600 & 140\\
12/30/12 & 0.59 & 7000 & 0 & 1100 & 36 & 600 & 80\\
1/5/13 & 0.97 & 7000 & 0.06 & 1650 & 11 & 600 & 260\\
1/8/13 & 1.16 & 7000 & 0.02 & 1150 & 28 & 600 & 90
\enddata
\tablecomments{Blackbody temperatures are given in $K$, solid angles are
given in units of the stellar solid angle assuming a radius of 1.5 R$_{\odot}$
and distance of 196 pc.}
\label{model_param}
\end{deluxetable}

\subsubsection{Emission lines}

DQ Tau exhibits an emission line spectrum that is fairly typical for
low to moderate accretors.  Table~\ref{emission_fluxes} lists the measured
line fluxes for the most prominent lines, including the Ca II triplet,
He I $\lambda1.08 \mu$m, and H I Paschen and Brackett lines.  This is not a
complete list of detections; upper Brackett lines in H band appear in some
epochs, but are severely blended with the copious photospheric absorption
lines in that range.  We also detect CO fundamental emission in M band
(lines of the P and R branches) at all epochs; however, we cannot reliably
measure these because of both residual telluric absorption and
blending of the individual lines themselves.

\begin{figure}[H]
\plotone{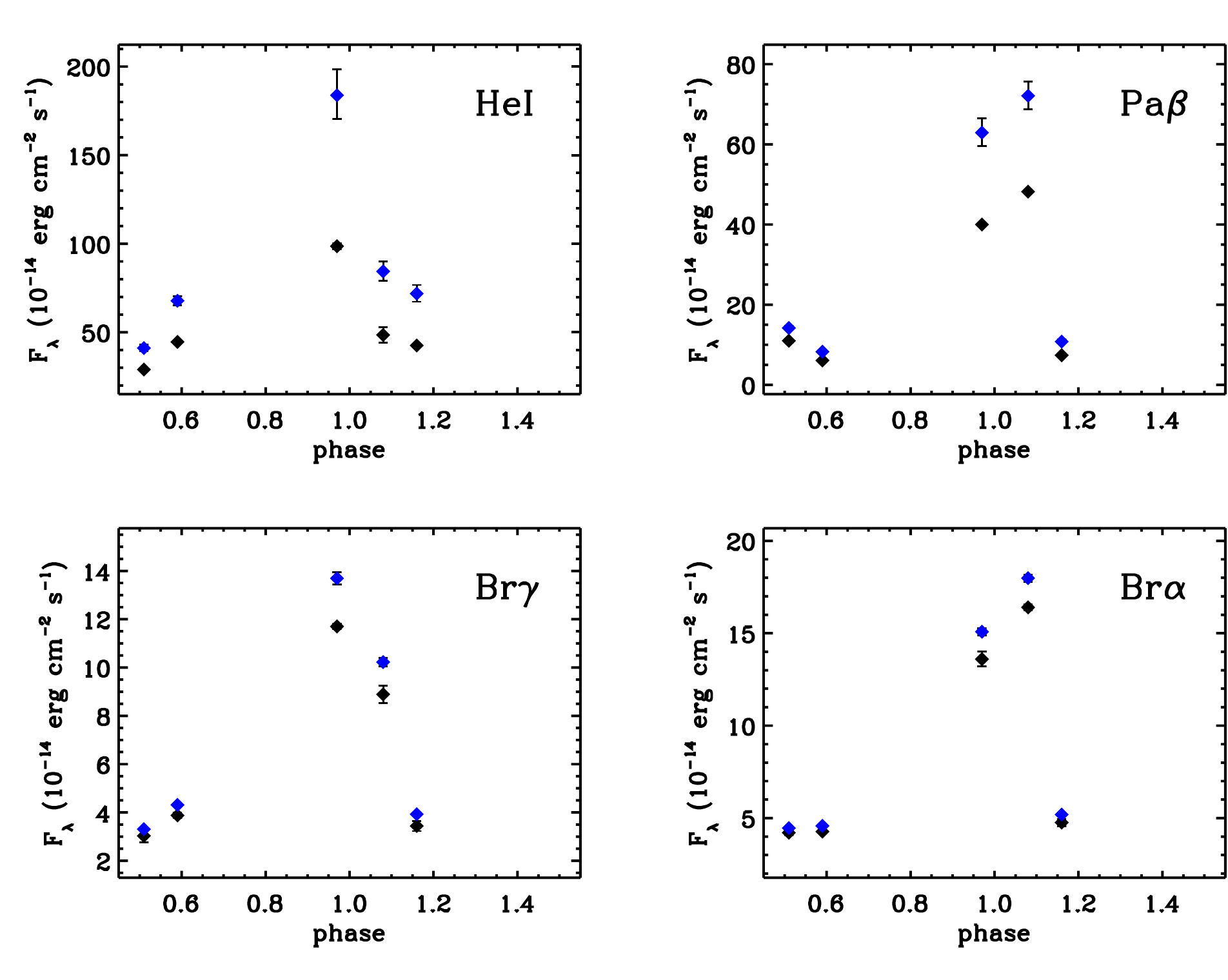}
\caption{Line flux vs. binary orbital phase for the four lines indicated.
Black and blue diamonds represent observed and dereddened fluxes, respectively.
The error bars for the dereddened fluxes include the uncertainty in
the reddening value derived from the spectral fits.
\label{emission_lines}}
\end{figure}

\begin{figure}[H]
\includegraphics[scale=0.7]{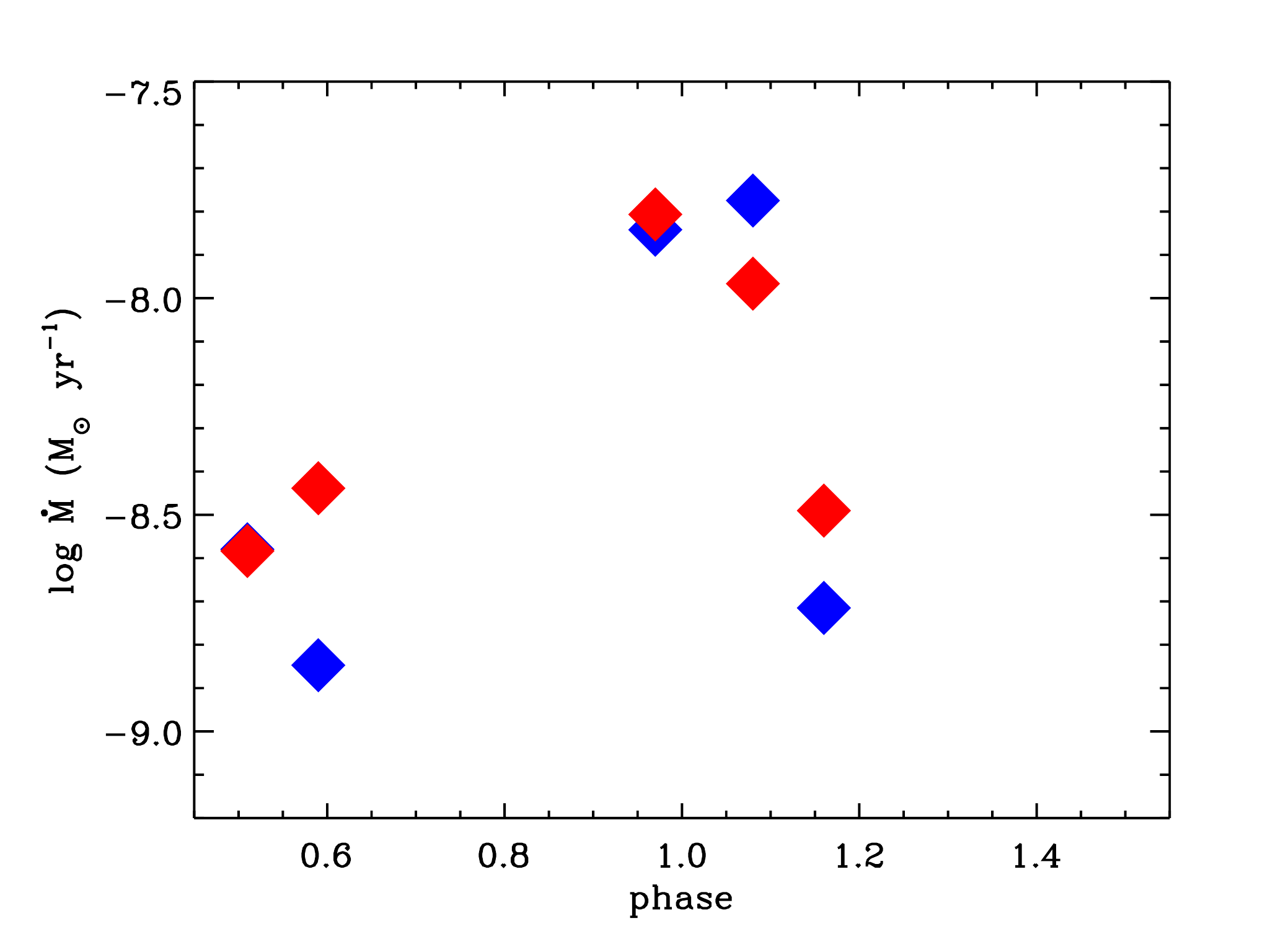}
\caption{Mass accretion rate vs. binary orbital phase. Blue and red diamonds
represent values derived from the measured Paschen $\beta$ and Brackett
$\gamma$ line fluxes, respectively.
\label{mdot}}
\end{figure}

All of the measured line emission varies significantly with time,
by a factor of 5 or more in some cases.
The strongest lines are clearly correlated with orbital phase
(Fig.~\ref{emission_lines}), with the flux peaking at or just before periastron
passages, similar to the optical photometry.  These lines are known
tracers of accretion activity in TTSs, and their observed variation
is fully consistent with the prior evidence for phase-dependent, or
``pulsed", accretion in the DQ Tau system.  Using previously-calibrated
relations between line luminosity and accretion luminosity
\citep[e.g.][]{1998AJ....116.2965M},
we estimated mass accretion rates from the observed
Pa$\beta$ and Br$\gamma$ line fluxes.  To do this, we applied extinction
corrections based on the simultaneous A$_V$ measurements from the veiling
analysis, and adopted the most recent determinations of the binary
stellar mass \citep[M$_* \sim 0.6$ M$_{\odot}$,
dividing in half the total mass from][]{2016ApJ...818..156C}
to convert from accretion luminosity to mass accretion rate.
The results are shown in Figure~\ref{mdot};
the accretion rate varies by roughly an order of magnitude
from $\sim 10^{-9}$ to $\sim 10^{-8} \; \msunyr$, strongly correlating
with orbital phase.  Our measurements are consistent with previous studies;
Bary14 (also based on NIR lines)
and \citet{2017ApJ...835....8T} (based on U band photometry), both with far
more epochs, also showed a strong correlation between accretion luminosity
and/or rate and orbital phase,
although both also showed some exceptions with increased activity
far from periastron.

\begin{deluxetable}{lccccc}
\tabletypesize{\small}
\tablewidth{0pt}
\tablecaption{DQ Tau emission line fluxes}
\tablehead{
\colhead{line} &
\colhead{12/13/12} &
\colhead{12/22/12} & 
\colhead{12/30/12} &
\colhead{1/5/13} &
\colhead{1/8/13}}
\startdata
O I $\lambda$8446 & 4.91 0.70 & 13.5 0.30 & 5.01 0.52 & 11.9 0.4 & 4.84 0.42\\
Ca II $\lambda$8498 & 2.51 0.17 & 7.50 0.19 & 2.58 0.75 & 9.74 0.33 & 2.34 0.80\\
Ca II $\lambda$8542 & 7.50 0.78 & 7.67 0.30 & 7.88 0.31 & 7.51 0.31 & 7.82 0.08\\
Pa11 & 1.30 1.30 & $^a$ & 1.0 1.0 & 2.36 0.27 & 1.8 1.8\\
Ca II $\lambda$8662 & 3.17 0.15 & 7.62 0.14 & 5.46 0.12 & 8.45 0.12 & 5.38 0.14\\
Pa 10 & $^a$ & $^a$ & $^a$ & $^a$ & $^a$\\
Pa 9 & 1.50 1.50 & 3.94 0.38 & 1.4 1.4 & 6.06 0.34 & 0.9 0.9\\
Pa 8 & $^a$ & $^a$ & $^a$ & $^a$ & $^a$\\
Pa 7 & 1.00 1.00 & 4.52 0.12 & 2.4 2.4 & 3.63 0.27 & 0.8 0.8\\
Pa 6 & 1.40 1.40 & 7.89 0.21 & 1.2 1.2 & 8.60 0.32 & 1.04 0.26\\
Pa 5 & 4.34 0.29 & 15.23 0.65 & 2.5 2.5 & 14.1 0.5 & 2.90 0.15\\
Pa$\delta$ & 2.47 0.53 & 18.3 0.15 & 5.6 5.6 & 15.8 0.2 & 3.5 3.5\\
He I & 29.0 0.70 & 48.5 4.30 & 44.6 0.5 & 98.6 1.7 & 42.6 0.6\\
Pa$\gamma$ & 6.85 1.26 & 30.2 0.80 & 4.74 0.47 & 30.2 0.3 & 4.75 0.23\\
O I $\lambda$1.13 & 2.88 0.15 & 9.78 0.28 & $^a$ & 9.98 0.31 & 7.0 7.0\\
Pa$\beta$ & 11.0 0.44 & 48.2 0.22 & 6.09 0.14 & 40.0 0.3 & 7.38 0.19\\
Br$\gamma$ & 3.03 0.27 & 8.89 0.36 & 3.88 0.11 & 11.7 0.1 & 3.44 0.21\\
Pf$\gamma$ & 2.1 2.1 & 3.68 0.27 & 1.5 1.5 & 3.25 0.17 & 0.66 0.18\\
Br$\alpha$ & 4.21 0.15 & 16.4 0.17 & 4.27 0.04 & 13.6 0.4 & 4.76 0.19\\
\enddata
\tablecomments{ In units of 10$^{-14}$ erg cm$^{-2}$ s$^{-1}$.
Identical flux and uncertainty values indicate upper limits.\\
$^a$ \, Blended with nearby stellar or uncorrected telluric absorption.}
\label{emission_fluxes}
\end{deluxetable}

\begin{figure}[H]
\includegraphics[scale=0.7]{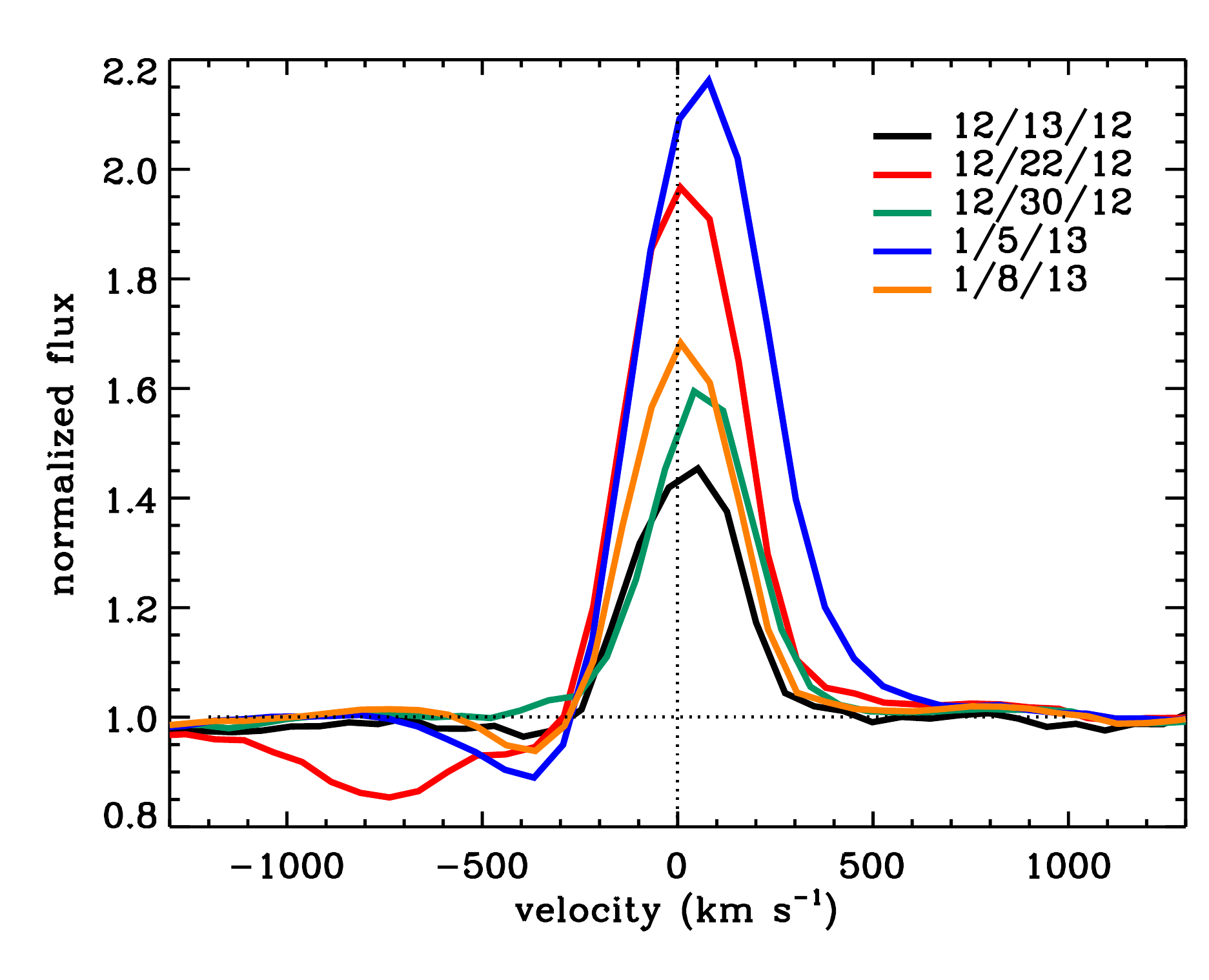}
\caption{The He I 1.083 $\mu$m line from the five epochs of SpeX spectra.
Relative shifts between the spectra introduced by uncertainties in
the wavelength calibration have been removed by comparing the velocities
of nearby photospheric absorption lines, arbitrarily using the 12/13/2012
spectrum as the baseline.  Given the range of these relative shifts,
we estimate that the absolute velocity scale is accurate
to $\sim 30$ km s$^{-1}$.  The velocity scale was further
corrected for the systemic radial velocity of 22 km s$^{-1}$.
Most of the individual emission or absorption components are likely
not resolved.
\label{he1}}
\end{figure}

The He I line at 1.083 $\mu$m is a particularly interesting diagnostic
of both accretion and outflow, as it often shows both blueshifted and
redshifted absorption superposed on the emission profile
\citep{2003ApJ...599L..41E, 2006ApJ...646..319E}.
This line is robustly detected in
emission in all of our spectra, with blueshifted absorption components
seen in three of the five epochs (Fig.~\ref{he1}). 
\citet{2006ApJ...646..319E} observed DQ Tau at high spectral resolution
at an orbital phase of $\sim 1.3$, and their profile
is consistent with the strength and shape of our observations
at quiescent epochs.  Our spectral resolution is too low to detect
the blueshifted absorption they observed at velocities around -200 km s$^{-1}$.
We detect a much more strongly blueshifted absorption component at
$\sim 400$ km s$^{-1}$ at the three epochs closest to a periastron passage,
which is significantly more blueshifted than seen in any of the T Tauri
line profiles shown by \citet{2006ApJ...646..319E}.
Bary14 also reported on variations of the He I blueshifted
absorption component in DQ Tau as a function of orbital phase,
though they do not remark on velocities (and many of their spectra
have insufficient resolution to reliably detect the absorption components).
The spectrum we observed on 12/22/2012 (orbital phase 1.08)
exhibits a second, stronger absorption component centered at roughly
{\it 750} km s$^{-1}$.  Such a velocity is virtually unprecedented
for any tracer of gas motions observed around any T Tauri star,
and is far above the escape velocity of the system.
This component does not appear to be a spurious artifact since
it is clearly seen at the same velocity and depth in each
of the 12 individual exposures
that were averaged into the final spectral extraction.
However, no absorption or emission at comparable velocities is seen
in any other observed line.  The appearance of this absorption
very close to a periastron passage, where the magnetospheres
of the two stars are likely to overlap, suggests a possible origin
in some kind of flare event caused by magnetic reconnection
(possibly analogous to a coronal mass ejection).
We cannot make an accurate estimate of the terminal velocity since
the absorption component is not spectrally resolved, but it could be
as high as $\sim 1000$ km s$^{-1}$.
Further observations near periastron
passage at higher spectral resolution are needed to determine whether
events of this type are common, and what the true origin might be.

\section{Discussion}

%
%
%
%
%

The combined results of our photometric and spectroscopic observations
show clear correlations of the accreting gas and hot dust with the binary
orbit in the DQ Tau system.  This behavior is broadly consistent with
the pulsed accretion scenario predicted by simulations.
Our spectroscopy indicates that the cavity inside the circumbinary disk,
if present, is never completely clear
of dust; the minimum dust temperatures we measure ($\sim 1100$ K)
around apastron orbital phases are much higher than the equilibrium
temperature at the location of the disk edge expected by simulations
($\sim 600$ K). However, those dust temperatures are also clearly lower
than the sublimation
temperature, which suggests that there is little if any dust near the
sublimation fronts of either star at those epochs.

\begin{figure}[H]
\includegraphics[scale=0.7]{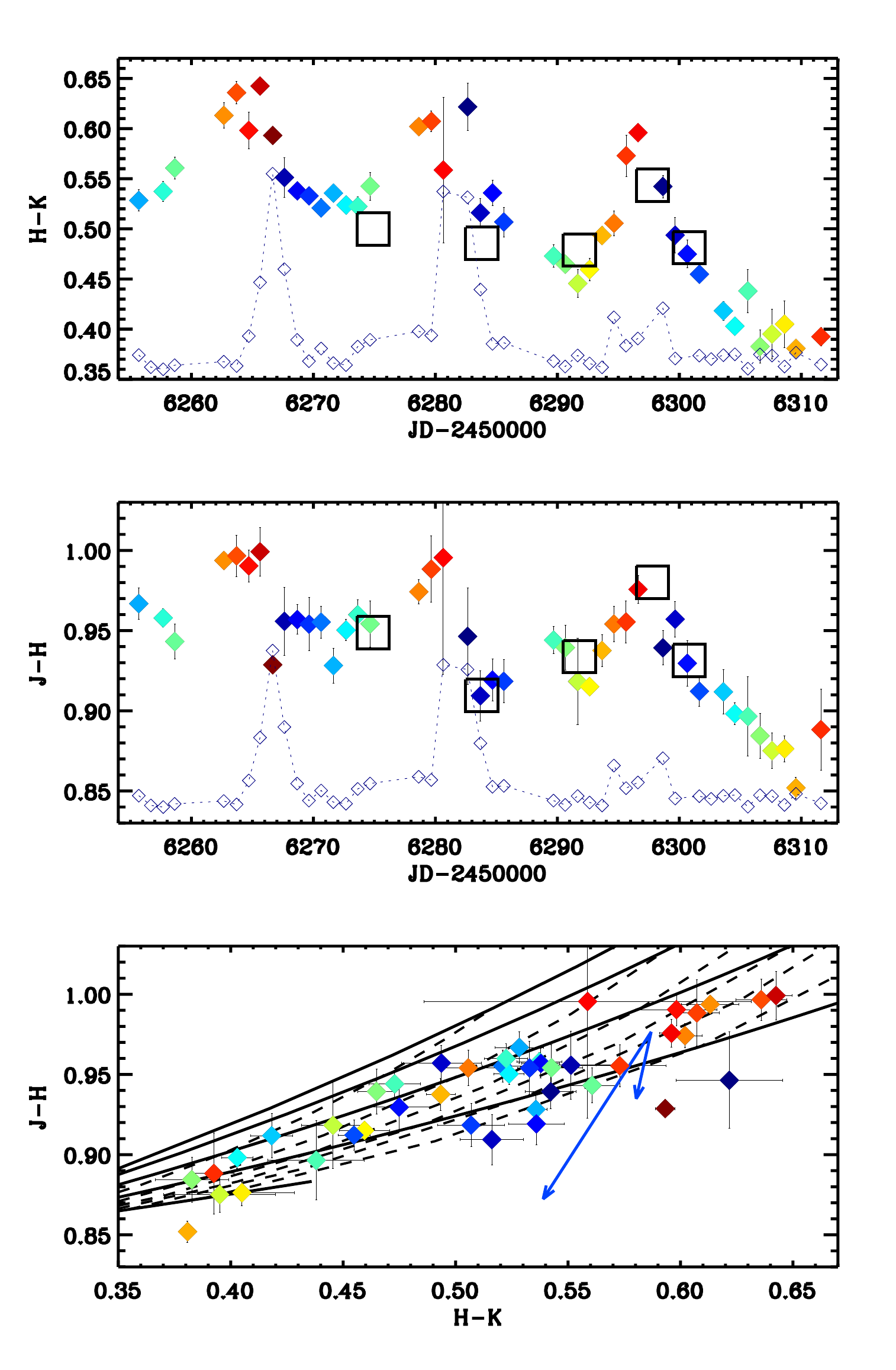}
\caption{Observed NIR colors for DQ Tau in a window around the times of
the spectroscopic observations, compared to stellar plus blackbody
models.  The color scheme of the observed points is based on orbital phase,
with green through red to brown representing the progression from
apastron (phase 0.5) to periastron (phase 1),  and black through blue
to green the progression from apastron to periastron.
(Upper panel) H$-$K color as a function
of time.  The light purple diamonds connected by a dotted line show
the scaled B band magnitudes for reference.  Large black squares represent
model colors for each spectroscopic observation derived by combining
the excess spectrum blackbody fits with the scaled stellar template spectrum.
(Middle panel) J$-$H color as a function of time.  The symbols are the same
as in the upper panel.
(Lower panel) H$-$K versus J$-$H colors, using the same phase-dependent
color scheme as in the upper panels.  The black lines represent
blackbody models using the same prescription as the models fit to the
excess spectra, but with no hot component and a constant extinction of
$A_V=1$.  Each solid line shows the locus of models for a single warm dust
component temperature (from top to bottom: 1700, 1500, 1300, 1100, 900 K).
Each dotted line connects models with the same warm component solid angle
(from left to right: 10, 20, 30, 40, 50, 70).  The blue arrows show
the shift in color that would result with an added hot component with
$T=7000$ K and $\Omega = 0.06$, for extinctions $A_V=1$ (left arrow)
and $A_V=1.8$ (right arrow).
\label{model_colors}}
\end{figure}

With only five epochs, the spectroscopy offers limited snapshots of the full
range of behavior of DQ Tau as traced by the photometric data.
Figure~\ref{model_colors} shows observed NIR colors 
for a subset of the photometric observations that span a time range
encompassing the spectroscopy. Corresponding colors derived from
the blackbody models of the excess spectra, with a scaled photospheric
template spectrum added in, are also indicated;
these match the contemporaneous
photometry to within the relative flux accuracy of the spectroscopic data.  
Note that four of the spectroscopic epochs span a very small range of
NIR colors, and the reddest epoch (1/5/2013)
is still somewhat bluer than the peaks in the H$-$K light curve.
We calculated
additional sets of blackbody models to see what range of parameters
would be needed to explain the maxima and minima of the NIR colors.
One of these sets, using the minimum measured extinction and assuming
no hot component, is shown in the lower panel of Fig.~\ref{model_colors}.
Although there is degeneracy between
the effects of extinction and hot accretion emission,
we can make some general conclusions. One, the rise of the NIR excess prior
to the onset of an accretion pulse is primarily explained by a significant
increase in the dust emission solid angle (by as much as a factor of 3).
An increase in the dust temperature may also occur, but the photometric
accuracy prevents a definitive constraint
in the absence of simultaneous spectroscopy.
The rise cannot be due to a significant increase in the extinction alone,
otherwise the $J-H$ color would be larger than observed.
Two, the highest dust temperatures, corresponding to the expected
sublimation limit, and the highest extinctions can only occur once
the accretion pulse is in progress; the blue excess serves to cancel out
the reddening effect at shorter NIR wavelengths, effectively putting a cap
on the $J-H$ colors.  At epochs with the strongest accretion emission,
there is a pronounced blueing of the $J-H$ color.  Near the end of a pulse,
both the $J-H$ and the $H-K$ colors drop, signaling a decline in both
the dust temperature and emission solid angle. Three, the warm dust
component exhibits a varying minimum temperature between periods of
quiescence, which may be related to the 75-day period possibly present
in the NIR photometry.  The minimum temperature may get as low
as $\sim$900 K, still warmer than the expected temperature of
the inner edge of the circumbinary disk.

\epsscale{1}

\begin{figure}[H]
\centering
\subfloat{%
  \raisebox{-0.5\height}{\includegraphics[width=0.5\columnwidth]{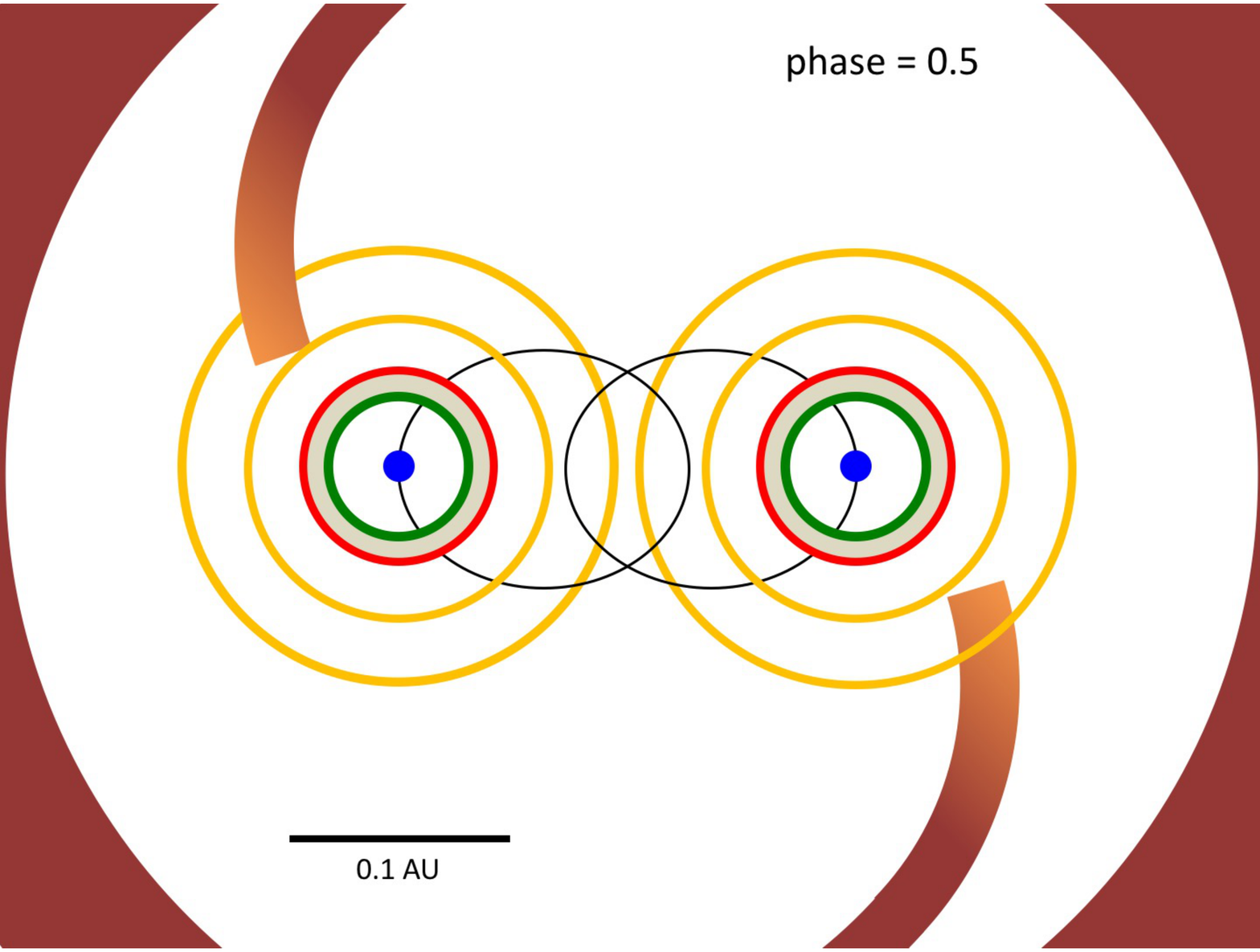}}%
}\qquad
\subfloat{%
  \raisebox{-0.5\height}{\includegraphics[width=0.4\columnwidth]{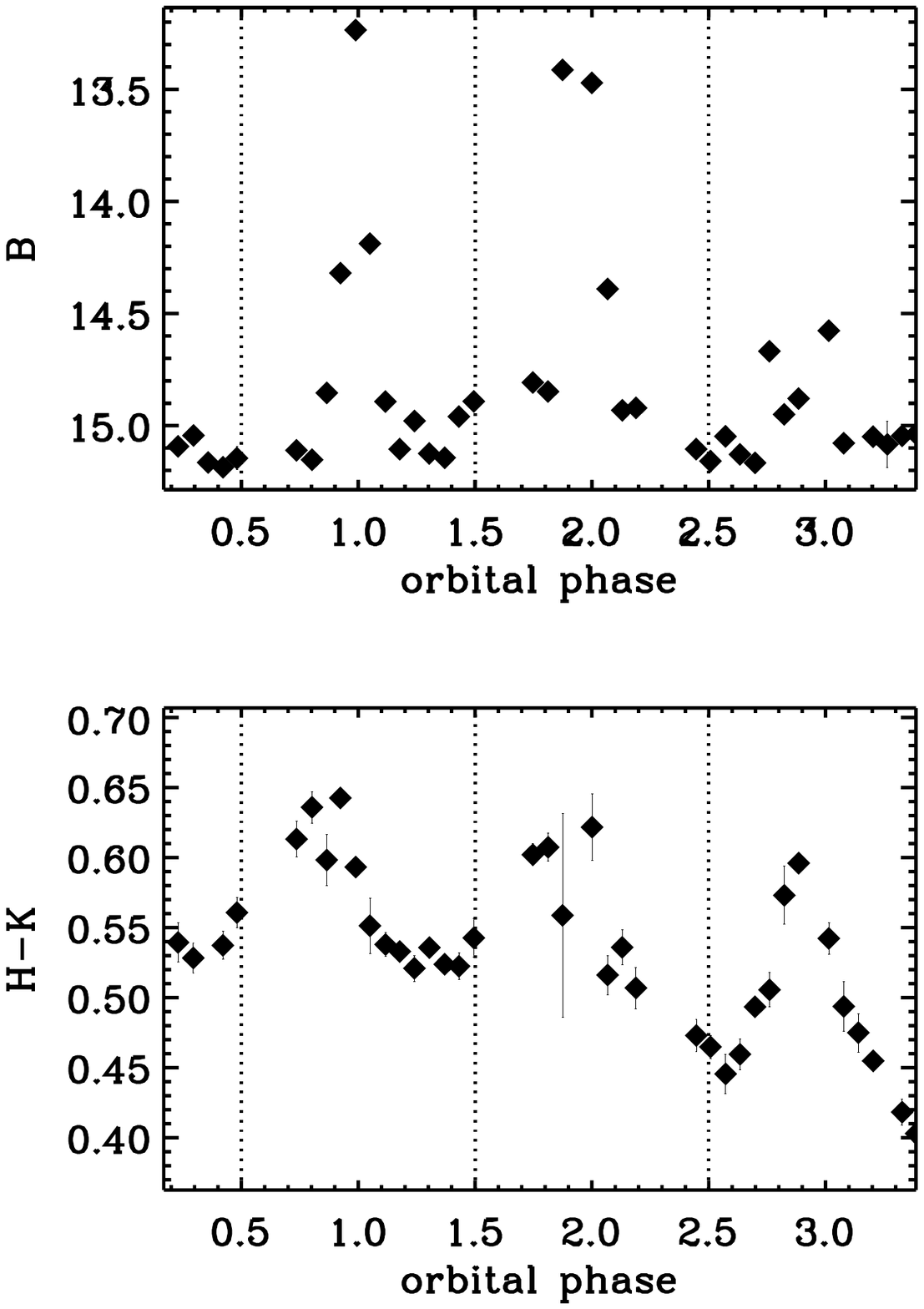}}%
}
\caption{(Left) A schematic of the DQ Tau system at apastron orbital phase.
The position of each star in its orbit is indicated by the blue dots,
and the stellar orbits are represented with black ellipses.
The corotation radius around each star, as estimated from the observed stellar
rotation period (Table~\ref{params}), is indicated with the green circles.
The red circles represent the theoretical dust sublimation radius around
each star, calculated assuming T $=$ 1650 K and negligible accretion luminosity.
The yellow circles show the approximate location of dust with an equilibrium
temperature of 1100 and 1300 K, again assuming negligible accretion luminosity.
The inner edge of the putative circumbinary disk and accretion streams
are shown in orange/red, and dust-free accreting circumstellar gas is shown
in gray.  All sizes are to scale.
(Right) A selection of the B band and H-K color light curves, with the
orbital phases corresponding to the schematic indicated with dotted
vertical lines.
\label{cartoon_0.5}}
\end{figure}

\begin{figure}[H]
\centering
\subfloat{%
  \raisebox{-0.5\height}{\includegraphics[width=0.5\columnwidth]{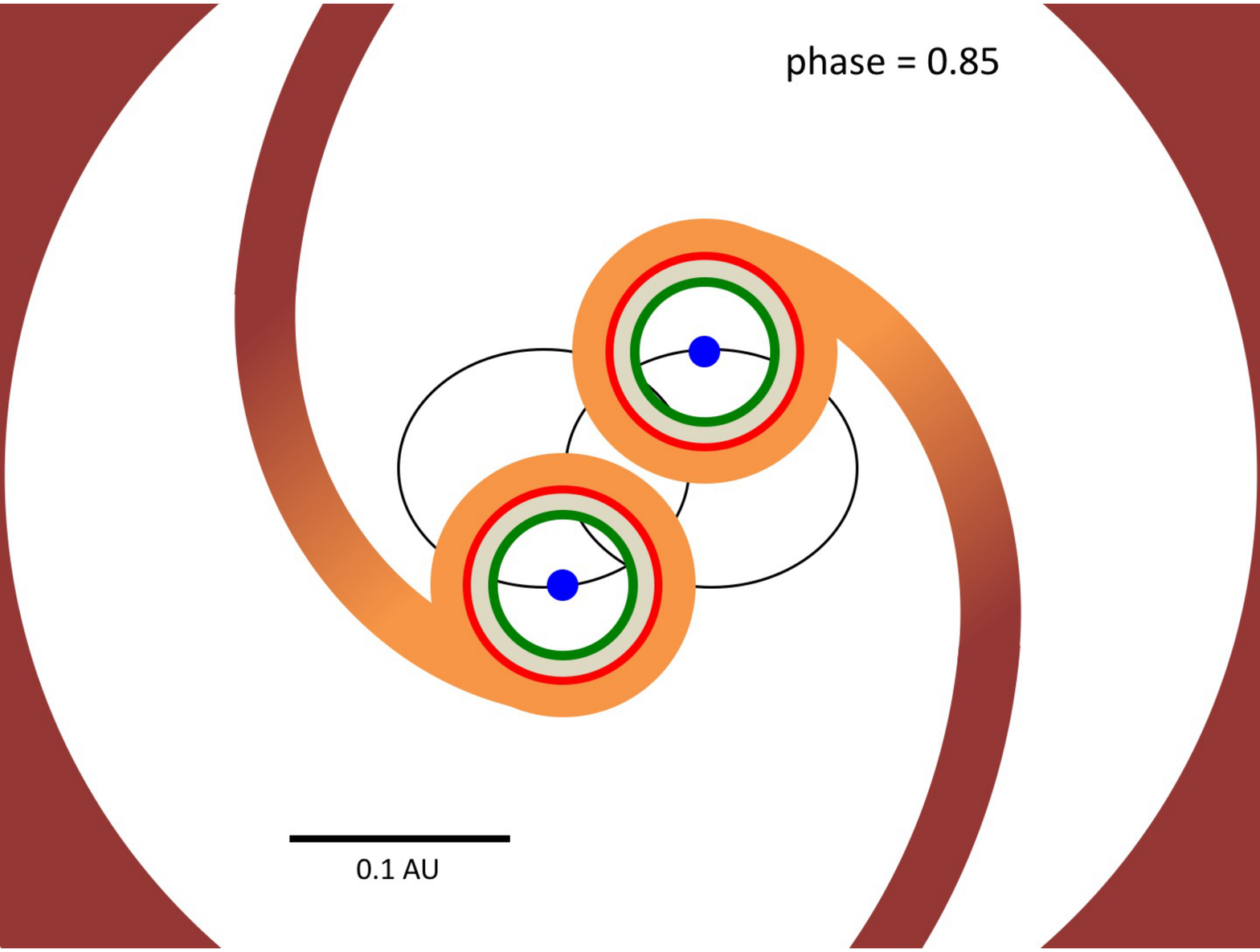}}%
}\qquad
\subfloat{%
  \raisebox{-0.5\height}{\includegraphics[width=0.4\columnwidth]{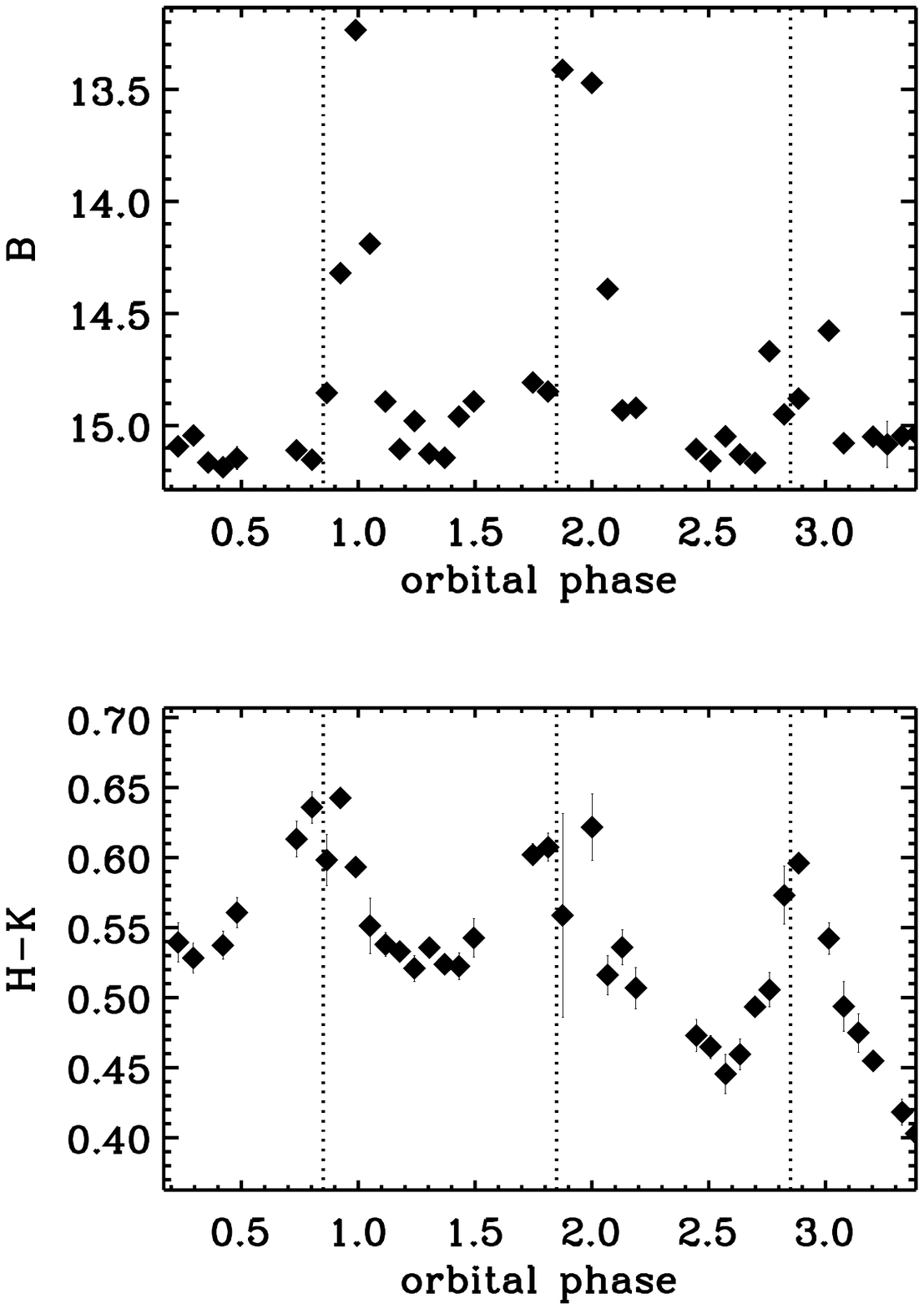}}%
}
\caption{Same as figure~\ref{cartoon_0.5}, for an orbital phase of 0.85.
Quasi-stable circumstellar disks begin to develop as material
is funneled inward, leading to a rise in the warm dust emitting area
and characteristic temperature.
\label{cartoon_0.85}}
\end{figure}

\begin{figure}[H]
\centering
\subfloat{%
  \raisebox{-0.5\height}{\includegraphics[width=0.5\columnwidth]{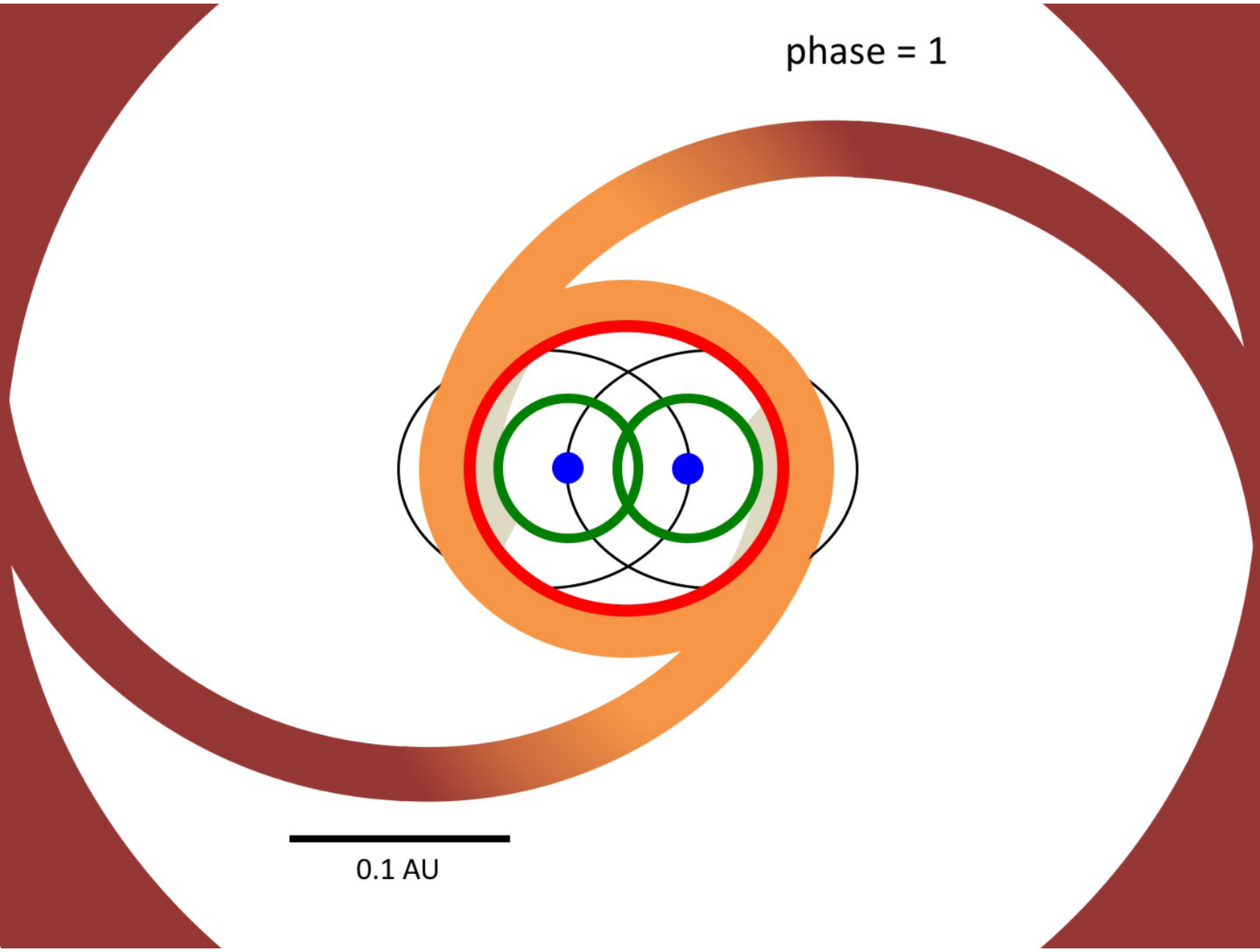}}%
}\qquad
\subfloat{%
  \raisebox{-0.5\height}{\includegraphics[width=0.4\columnwidth]{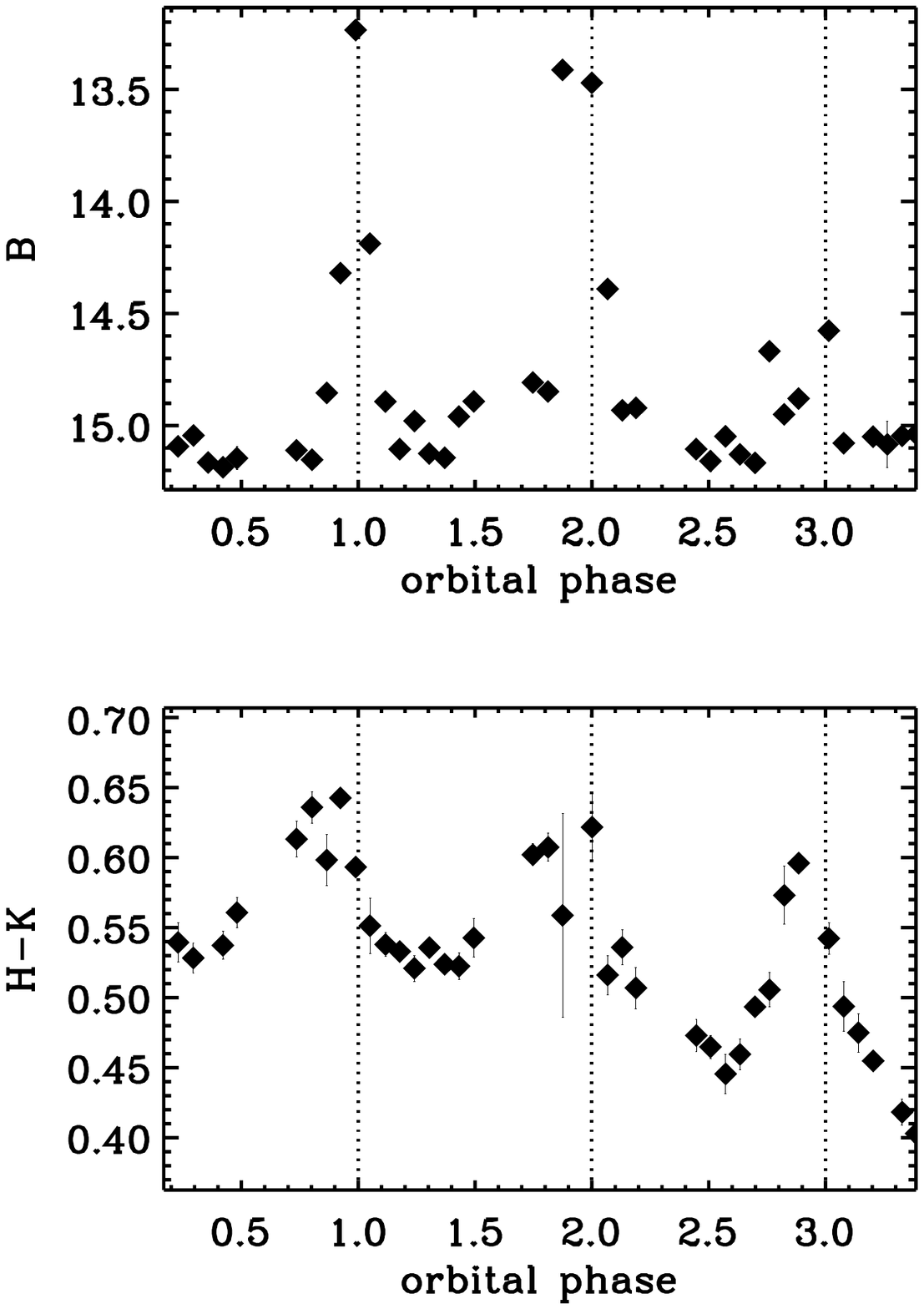}}%
}
\caption{Same as figure~\ref{cartoon_0.5}, for periastron orbital phase.
During closest approach the circumstellar dust structures coalesce
around both stars, and a burst of accretion occurs as the inner
circumstellar gas is disrupted and rapidly falls onto the central stars.
The sublimation front indicated here was calculated assuming
the combined luminosity of both stars and a contribution
from the accretion luminosity, given as the maximum measured level
from our emission line observations.  
\label{cartoon_1}}
\end{figure}

\begin{figure}[H]
\centering
\subfloat{%
  \raisebox{-0.5\height}{\includegraphics[width=0.5\columnwidth]{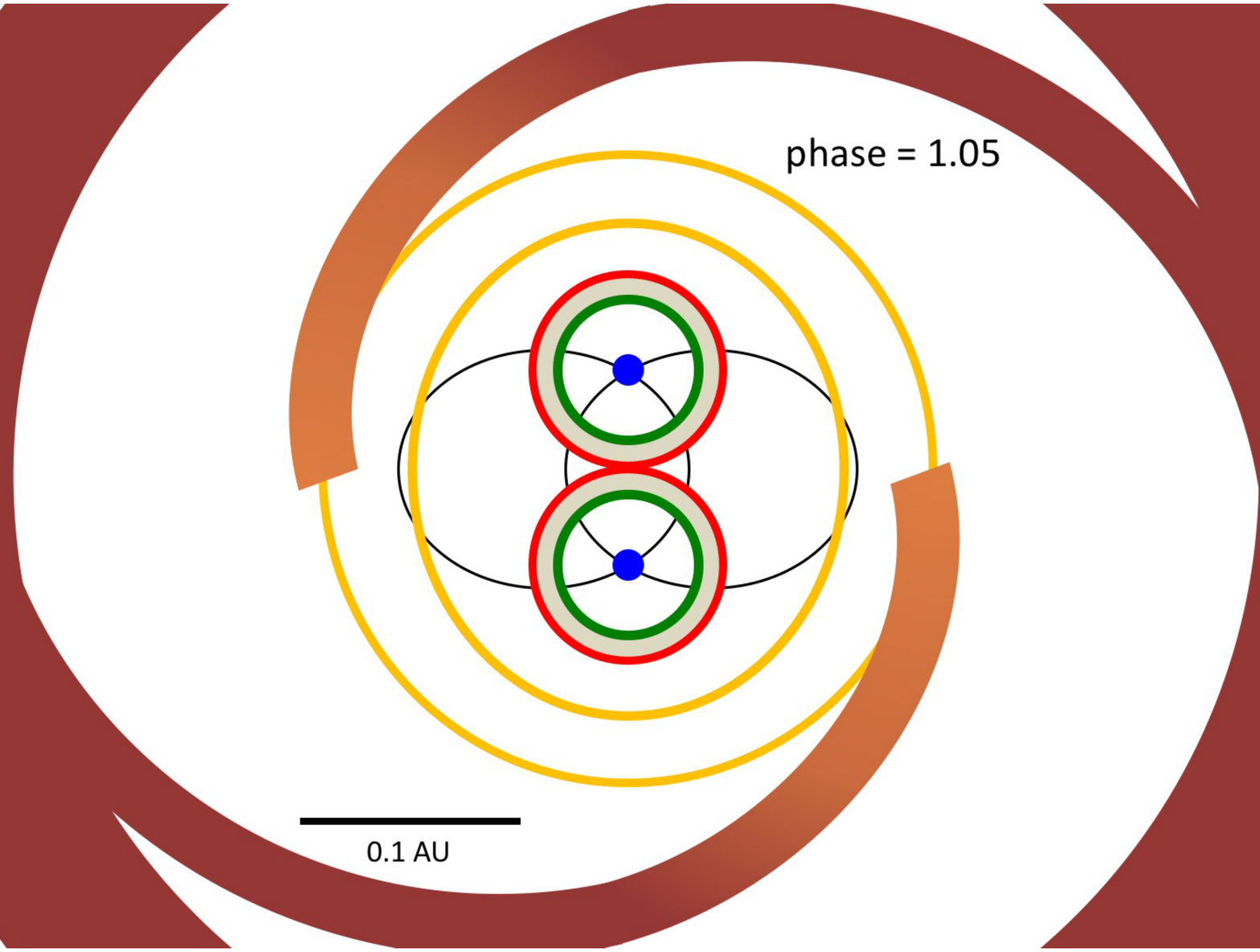}}%
}\qquad
\subfloat{%
  \raisebox{-0.5\height}{\includegraphics[width=0.4\columnwidth]{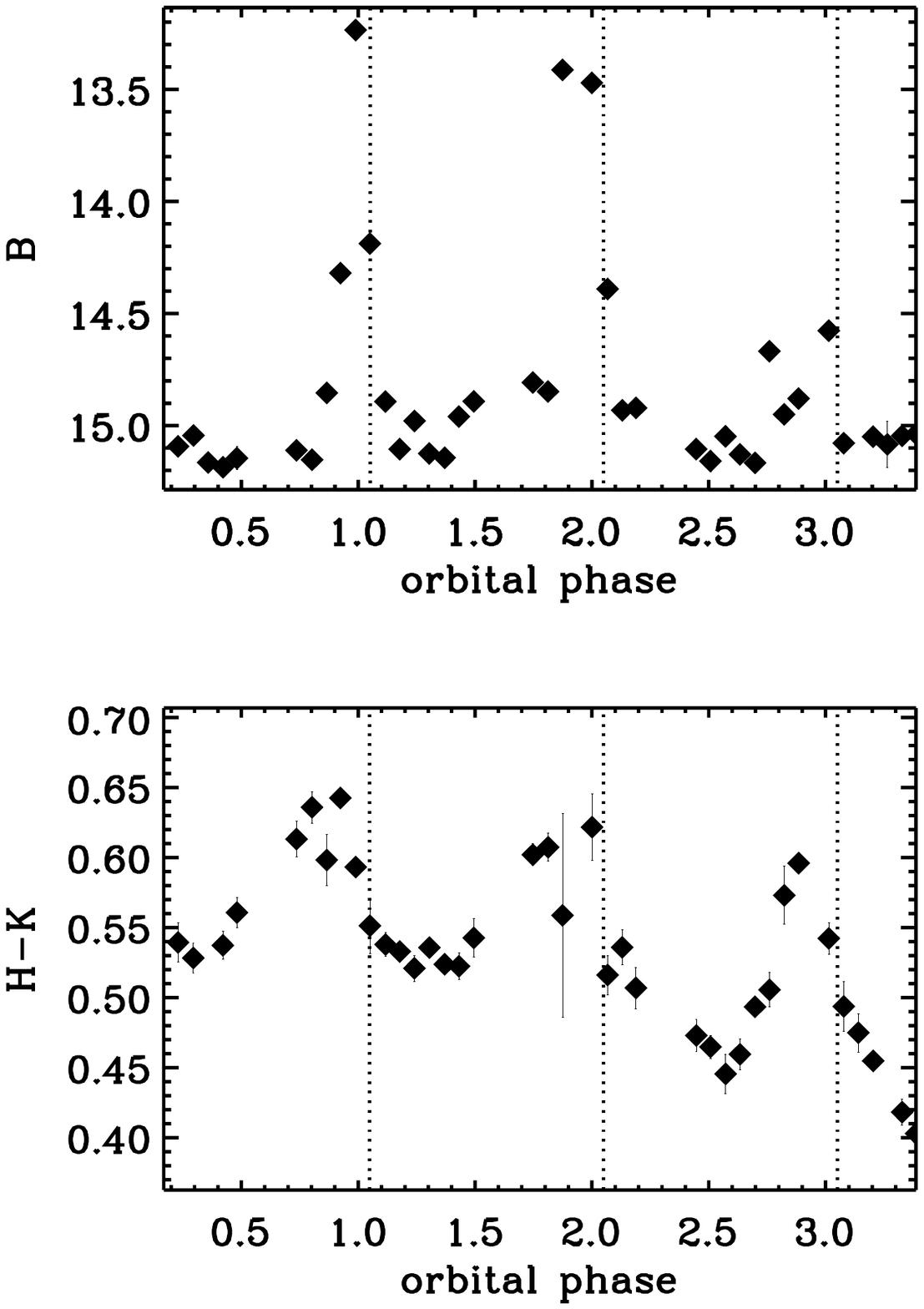}}%
}
\caption{Same as figure~\ref{cartoon_0.5}, for an orbital phase of 1.05.
As the stars move further apart, the circumstellar dust structure is disrupted,
leading to a drop in the infrared excess strength and characteristic temperature,
as well as the accretion rate onto the stars.
The two yellow circles mark approximate locations for dust at
T$=$1100 and 1300 K, assuming the combined stellar luminosity and negligible
accretion luminosity.
\label{cartoon_1.05}}
\end{figure}

Using prior simulations as a guide, we propose the following scenario as
a general framework for understanding the observations, using schematics
of the system geometry as shown in Figures~\ref{cartoon_0.5}-\ref{cartoon_1.05}.
Note that these present a highly idealized picture of the complex dynamical
environment (for example, the circumbinary disk inner edge is likely not
sharp and well-defined, and the material inside is not all strictly
confined to narrow streams), but are intended to provide a rough guide
to the scale of various physical regions of interest.
To start, at apastron orbital phase,
accretion streams falling from the inner edge of the circumbinary disk
begin to approach each star (Fig.~\ref{cartoon_0.5}).  Most of the material has
not yet reached the sublimation radius around either star, hence the apparent
dust temperature should be lower than the typical sublimation point.
There is still gas left around one or both stars,
located mostly between the sublimation front and corotation radius,
slowly accreting onto the star(s) at a quiescent level after
having been cut off from the previous orbit's streams.  
As the stars come closer together in their orbits, the accretion streams
fall inwards, increasing the amount of gas and dust inside the cavity,
and begin feeding temporary circumstellar
structures (Fig.~\ref{cartoon_0.85}).  The dust in the streams reaches
the sublimation fronts, and increases the gas supply to the accretion flows
onto the stars, resulting in an increasing accretion rate.  The increase
in the amount of circumstellar material, and possibly its scale height,
leads to an increase in the NIR emission (and possibly the stellar extinction
if the material is sufficiently stirred out of the disk plane).
As the stars continue to draw closer, the circumstellar dust structures
begin to interact.  Near periastron phase, there is a single sublimation
front around both stars, the inner gas disks become disrupted as
the corotation radii slightly overlap, and the stellar magnetospheres
may also interact (Fig.~\ref{cartoon_1}).  This produces a spike in the stellar
accretion rate.  Finally, as the stars begin to separate, the circumstellar
dust structures are torn apart, and the accretion streams become disconnected,
leading to a drop in the NIR excess flux and characteristic temperature,
and a decrease in the stellar accretion rate (Fig.~\ref{cartoon_1.05}).

This generalized picture does not explain all observed
characteristics.  For one, the accretion of gas onto the binary never
completely stops, which means there must always be some reservoir of
material located near the corotation radius of one or both stars
\citep[the viscous timescale at that location is of order 100 yr;][]
{1998ApJ...495..385H}.
Some material also likely remains around
the sublimation radius, which should lead to some hot dust emission;
if so, it may be that such emission is too small to detect along with
the cooler dust emission that is dominant during the quiescent epochs
(from the excess spectra, we estimate a conservative limit on the solid
angle of a component at the sublimation temperature to be $<10$\%
of the dominant cooler component).
Another difficulty lies in explaining the very rapid decline in dust
emission and temperature immediately after periastron orbital phase.
It seems unlikely that all dust near the sublimation front
gets dynamically disrupted in one or two days; again, there may be
an emission component that is too small to reliably identify
in the excess spectrum.
Finally, not all of the monitored binary orbital cycles exhibit such a clear-cut
pulse signature in either the optical or NIR wavelength range.
A few cycles show very complicated photometric behavior, with multiple
peaks in both the optical and NIR.  These may indicate a breakdown of
the regular two-armed accretion stream infall, perhaps by turbulent clumps
accreted from the circumbinary disk.
In any case, more detailed simulations, and calculation of observables
from them, are needed to clarify specific predictions from
the pulsed accretion theory.

In cycles with a strong NIR peak, the observed $J-H$ colors rule out
the presence of a significant amount of dust near the sublimation front
($T \sim 1600$ K) prior to the start of the stellar accretion pulse.
The fact that we measured a 1650 K component in one spectrum during
an accretion pulse suggests a connection between the sudden increase in
dust temperature and the pulse itself.
This could simply be a result of the sudden expansion of the sublimation
radius to encompass both stars when they move sufficiently close together
(as shown in Fig.~\ref{cartoon_1}).  There may also be a concurrent
``puffing up" of the material near the sublimation front as irradiation
heating from the accretion luminosity increases.
Enhanced outflow related to the accretion pulse may also
entrain more dust, raising the effective solid angle of the hottest
dust \citep{2012ApJ...758..100B}.  Another possibility is the formation of
shocks as the accretion streams collide with other material in the cavity,
as seen in simulations \citep[e.g.][]{2016ApJ...827...43M} and observed
via H$_2$ emission in the wide binary GG Tau \citep{2012ApJ...754...72B};
such shocks could potentially heat the circumstellar dust
and increase its scale height.
Our simplistic blackbody fits infer a very large solid angle for the warm
dust component near NIR photometric peaks prior
to an accretion pulse, with $\Omega \sim 50 \, \Omega_*$
or more for $T_{warm} \geq 1200$ K.  This is much larger than inferred
for single T Tauri stars (McClure13), and may be indicative of a higher
scale height for the temporary circumstellar dust structures
that form in the DQ Tau system close to periastron orbital phase,
as might be expected from dynamical stirring of the inner disk
material by the binary motion.
However, given the very different geometry, a detailed comparison
with the standard accretion disk model requires analysis of circumbinary
disk simulations, which is beyond the scope of this work.

Several previous investigations of warm material in the DQ Tau system
support our findings.  \citet{2001ApJ...551..454C} characterized CO fundamental
emission line profiles, finding a CO excitation temperature of 1200 K.
Assuming a simple Keplerian disk model, they estimated an emitting region
in the range $\leq 0.1$ to $\sim 0.5$ AU, which is at least partly co-located with
the warm dust component.
\citet{2009ApJ...696L.111B} used Keck NIR interferometry
to resolve K-band emission in the system, placing it at roughly 0.15-0.2 AU,
which again is consistent with our inferred location of the warm component.

In contrast, \citet{2018ApJ...862...44K} recently presented a somewhat
different characterization of the inner disk region, based on contemporaneous
observations from the {\it Kepler K2} mission and the {\it Spitzer} Space
Telescope.  Their {\it Spitzer} data included 3.6 and 4.5 $\mu$m photometry
obtained at a roughly daily cadence spanning slightly less than one
orbital period, and covering most of one accretion pulse event;
the light curves in both bands correlated with the {\it Kepler} optical data,
exhibiting an increase in flux between orbital phases of about 0.8 - 1,
and peaking a few days before periastron.  There is no evidence of
a lag between the infrared and optical data, unlike what we have found
for most pulse events at shorter infrared wavelengths.  However, the
{\it Spitzer} observations began when the optical pulse was already underway,
so it is possible the onset of the infrared flux increase may have occurred
before that of the optical flux.  We also note that their measured $[3.6]-[4.5]$
colors get progressively bluer before periastron, and redder thereafter,
which is qualitatively consistent with the dust temperature changes
we have inferred from our data.

Based on the {\it Spitzer} colors, Kospal et al. inferred a characteristic
dust temperature of 917 K near periastron passage, declining to 825 K
a few days later.  These temperatures are significantly cooler than
our spectroscopically-derived values; we believe their estimates
to be in error because they are based on single-temperature blackbody fits
to two photometric points spanning a limited wavelength range.  As we have
shown, the infrared excess emission shape can only be explained with
a range of dust temperatures (even the two dust components we adopted
are likely overly simplistic compared to reality), and the {\it Spitzer}
photometry by itself is relatively insensitive to any warmer dust component.
Since the {\it Spitzer} data coincided with an optical accretion pulse,
there must have been material closer to the stars than this location in
order to feed the stellar accretion flows, thus warmer dust
was almost certainly present.
Given the characteristic dust temperatures they inferred, Kospal et al.
estimated a physical location for the emitting material at about 0.13 AU.
Assuming this location corresponds to the inner edge of the circumbinary
disk, they interpreted the increase in characteristic dust temperature
near periastron as due to increased irradiation by the accretion pulse.
Our results suggest that changes in the location and geometry of
the emitting material are also important, if not dominant, in setting
the level and shape of the dust emission.  In addition, the warmer dust
temperatures we infer ($>1100$ K) indicate that most of this material
is unlikely to be associated with any stable circumbinary disk
since it is located in a region where the binary torques
are very strong.  Our data show ambiguous evidence of
the effects of variable irradiation on the warmest circumstellar material,
except perhaps in the case of the extremely large pulse we observed
near MJD-2456630. 
Nevertheless, further NIR spectroscopy tracing the onset and
growth of the NIR photometric peaks is required to better understand
the relative roles of dust heating and surface area.

As mentioned above, accretion pulses do occasionally occur at phases
other than periastron.  We observed several examples
of weak pulses a few days after periastron passage in the 2013-2014 season,
and one cycle with four roughly equally spaced peaks.
Other studies have also seen pulses occurring close to apastron phase
\citep[Bary14;][]{2017ApJ...835....8T, 2018ApJ...862...44K}.
The origin of these events is unclear;
they may represent stochastic accretion events related to clumpy
circumstellar material, as often seen in classical T Tauri disks
\citep{2014AJ....147...82C, 2014AJ....147...83S, 2016AJ....151...60S}.
\citet{2017ApJ...835....8T} suggested the possibility that the stars
may occasionally cross through
remnant streams from a previous cycle, as seen in some simulations.
Interestingly, the cycle we observed containing four optical peaks
occurred after the extremely large pulse near MJD-2456630, and also
exhibited the reddest $H-K$ baseline level, indicative of a significant
amount of residual material inside the circumbinary disk.

The correlation between the $H-K$ minima and periastron passages with
weak or no optical pulses suggests a link between pulse strength and
the amount of material dragged inward by the accretion streams.
The source of the 75-day periodicity may be a result of dynamical
processes in or near the inner edge of the circumbinary disk.  For example,
some simulations have shown that the disk cavity can become eccentric
and/or develop repeated azimuthal density enhancements
\citep{2012ApJ...749..118S, 2017MNRAS.466.1170M}. The former effect
is unlikely to lead to the observed variability since the eccentricity precesses
on a much longer timescale, of order years.
In the latter case, when a density ``lump" rotates into one of the points
from which the streams originate, it can lead
to an increase in the surface area and accretion rate in the stream.
If the 75-day periodicity is tracing orbital motion, the source would be
located at $\sim 0.37$ AU, very similar to the predicted location of
the lump at $\sim 3a$, which corresponds to $\sim 0.39$ AU in DQ Tau.
However, the simulations to date find that such density enhancements
form only in binaries with very low eccentricities, which is not the case
for the DQ Tau system.

Finally, our finding of an increase in the extinction near periastron phase
has implications for interpretation of optical light curves.
Some of the complicated structure seen at shorter timescales
\citep[e.g.][]{2017ApJ...835....8T, 2018ApJ...862...44K}
may be due at least in part to variations in the exinction
along the line of sight.  Because of the degeneracy between
extinction and accretion excess, it is impossible to reliably disentangle
the two effects with photometry alone.  Spectroscopy is essential
to breaking this degeneracy by using the method of veiling characterization
as a function of wavelength.

\acknowledgements
We acknowledge Steve Lubow, Jeff Bary, Ben Tofflemire, Bob Mathieu, and Bo Reipurth
for helpful discussions and encouragement.
J. M. extends a special thanks to the staff at IRTF, particularly Alan Tokunaga and John Rayner,
for their generous scheduling flexibility and peerless remote observing support.

\bibliographystyle{aasjournal}
\bibliography{references}

\end{document}